\documentclass[fleqn,usenatbib]{mnras}

\usepackage{newtxtext,newtxmath}

\usepackage[T1]{fontenc}

\DeclareRobustCommand{\VAN}[3]{#2}
\let\VANthebibliography\thebibliography
\def\thebibliography{\DeclareRobustCommand{\VAN}[3]{##3}\VANthebibliography}

\usepackage{graphicx}	
\usepackage{amsmath}	
\usepackage{mymacros}
\usepackage{tabularx}
\usepackage{threeparttable}
\usepackage{float}

\title[High-velocity X-ray Sources]{A comprehensive search for high-velocity X-ray sources: New compact object binary candidates in the \gaia\ era}

\author[Y. Zhao et al.]{Yue Zhao,$^{1}$\thanks{E-mail: Yue.Zhao@soton.ac.uk}
Poshak Gandhi,$^{1}$
Christian Knigge,$^{1}$
Phil Charles,$^{1,2,3}$
Daniel Stern,$^{4}$
Peter Boorman,$^{5}$
\newauthor
Pornisara Nuchvanichakul,$^{1}$
Cordelia Dashwood Brown,$^{1}$
and David A.H. Buckley$^{6,7}$
\\
$^{1}$School of Physics \& Astronomy, University of Southampton, Highfield, Southampton SO17 1BJ, UK\\
$^{2}$Astrophysics, Department of Physics, University of Oxford, Keble Road, Oxford OX1 3RH, UK\\
$^{3}$Department of Physics, University of the Free State, PO Box 339, Bloemfontein 9300, South Africa\\
$^{4}$Jet Propulsion Laboratory, California Institute of Technology, Pasadena, CA 91109, USA\\
$^{5}$Max-Planck-Institut f\"{u}r Extraterrestrische Physik, Gie{\ss}enbachstraße 1, 85748 Garching, Germany\\
$^{6}$South African Astronomical Observatory, PO Box 9, Observatory 7935, South Africa\\
$^{7}$Department of Astronomy, University of Cape Town, Private Bag X3, Rondebosch 7701, South Africa\\
}

\date{Accepted XXX. Received YYY; in original form ZZZ}

\pubyear{\the\year{}}

\begin{document}
\label{firstpage}
\pagerange{\pageref{firstpage}--\pageref{lastpage}}
\maketitle

\begin{abstract}
We perform a comprehensive search for high-velocity X-ray sources with large X-ray/optical flux ratios ($\fxfg$), identifying candidates for interacting black hole or neutron star binaries potentially accelerated by supernova natal kicks. We cross-match X-ray points sources from a variety of catalogues (\chandra, {\it XMM-Newton}, \swift\ and {\it eROSITA}) with \gaia\ DR3. Using \gaia\ coordinates, parallaxes, and proper motions, we compute peculiar velocities ($\vpec$) relative to Galactic disc rotation. Remaining agnostic about radial velocities (RVs), we vary RVs to find the minimum possible $\vpec$ values ($\vpecmin$). Uncertainties on $\vpecmin$ are estimated via Monte Carlo resampling, and we select X-ray sources that have $1\,\sigma$ lower limits on $\vpecmin\geq \vpecminlim\,\kms$ and high $\fxfg$ values. We show that this velocity threshold excludes most contaminants (e.g., cataclysmic variables and active binaries) while retaining a sensible fraction of compact object binaries, demonstrating that $\vpec$ could serve as an effective indicator for the presence of a neutron star or black hole companion. Our selection yields a sample of $\numhvxaftercleaning$ sources, from which we construct a gold sample of $\goldsamplesize$ sources that have relatively well-constrained astrometry and confident optical counterparts. Follow-up is necessary to confirm and characterise their high-energy emission, as well as a Galactic disc vs. halo origin.
\end{abstract}

\begin{keywords}
X-rays: binaries -- binary: close -- stars: black hole -- stars: neutron -- supernova: general
\end{keywords}


\section{Introduction}
\label{sec:introduction}
Compact object binaries (COBs), systems in which a neutron star (NS) or a black hole (BH) orbits a non-degenerate companion, have yielded critical insights into compact objects and their associated physics. NSs, for instance, serve as natural laboratories for studying matter under extreme densities, offering a unique opportunity to probe the physics of ultra-dense matter \citep[e.g.,][]{Lattimer12}. Observations of emission from regions of extreme gravity near these compact objects enable tests of relativistic effects \citep[e.g.,][]{Dvali13, Giddings14, Kleihaus11}. X-rays in these systems are produced by interactions between the compact object and its non-degenerate companion. Their emission traces high-energy processes such as accretion onto the compact object in X-ray binaries \citep[XRBs; e.g.,][]{Frank92} or interactions between relativistic winds and the companion's material in binary pulsars \citep[BPSRs; e.g.,][]{Bogovalov19}. Moreover, BHs and NSs have likely all experienced supernovae, making COBs a valuable probe for studying supernova mechanisms and energetics \citep[see e.g.,][for a review]{Janka12}.

The non-degenerate companion in a COB, also referred to as the optically luminous component, plays an important role in observing and characterising the ``dark" compact objects. By tracking the orbital motion of the luminous component, studies have constrained their binary orbits, and, in some cases, the mass of the compact object can be determined dynamically --- providing a definitive classification as either a BH or NS \citep[e.g.,][]{Thompson19, El-Badry23a}. The luminous component also serves as a tracer of the binary's systemic velocity, offering insights into the potential energetics of its birth supernova. It is broadly accepted that a supernova imparts an impetus to the BH or NS, referred to as a natal kick (NK), via ejection of baryonic matter \citep{Blaauw61} and/or asymmetric emission of neutrinos \citep{Chugai84, Dorofeev85, Arras99}. The binary's kinematic properties therefore provide an observational connection to the underlying physics of NKs.

Despite an ever-growing sample, our understanding of compact object demographics remains significantly incomplete. This is mainly due to the dearth of confirmed examples. There are only $\approx 30$ BHs that have been dynamically confirmed \citep{Corral-Santana16}, but the consensus is that our galaxy contains $\sim 10^8$ BHs \citep[e.g.,][]{Olejak20}; similarly for NSs, while over $\approx 3,000$ have been discovered as radio pulsars \citep{Manchester05}, this only represents a small ($\sim 10^{-5}$) fraction of the predicted population \citep[e.g.,][]{Faucher-Giguere06, Gulon14, Cieslar20}. Discovery of more COBs provides an important channel of enriching our sample of NSs and BHs and is particularly useful in constraining the related binary evolution models.

Systematic searches for new COBs have gained significant momentum since the {\it ROSAT} All-Sky Survey \citep[RASS;][]{Truemper82, Voges96, Voges99, Voges00, Boller16}. RASS surpassed previous surveys in both sensitivity and angular resolution \citep{Voges00}, and RASS sources have been cross-matched against optical catalogues to search for new XRBs \citep[e.g.,][]{Motch97}. Since the turn of the millennium, even fainter sources are detected by instruments with larger collecting areas, such as the {\it Chandra X-ray Observatory} (\chandra), the {\it XMM-Newton Telescope} (\xmm), and the {\it Neil Gehrels Swift Observatory} (\swift); however, these searches have been focused on restricted regions given the small fields of view of these instruments \citep[e.g.,][]{Jonker11, Bahramian21}. Some of these sources have been identified by optical/near-IR follow-up or cross-referencing with existing catalogues \citep[e.g.,][]{Nebot13, Shaw20}. The \gaia\ mission \citep{GaiaCollaborationandPrusti16} has made it possible to identify optical counterparts to X-ray sources on large scales. In fact, searches just based on \gaia\ astrometry \citep[e.g.,][]{Shahaf23} and/or photometric data \citep[e.g.,][]{Gomel23} have already identified many new candidates, including some non-interacting COBs (NICOBs) that have been dynamically confirmed \citep[e.g.,][]{El-Badry23a, GaiaCollaborationPanuzzo24}. Efforts have also been made combining non-single star information inferred from \gaia\ astrometric fits and archival X-ray catalogues for new XRB candidates \citep[e.g.,][]{Gandhi22}. Eventually, the eROSITA all-sky survey \citep[\erass;][]{Predehl21, Merloni24} will provide X-ray coverage of the full sky at unprecedented depths. The first data release has already revealed $930,203$ sources over the western Galactic hemisphere at an X-ray ($0.2-2.3\,\kev$) depth of $\gtrsim 10^{-14}\,\ergscm$ \citep{Merloni24}, and some new COB candidates have been discovered. For example, \citet{Zainab24} have followed-up one of the resulting XRB candidates, finding optical spectral features typical of high-mass XRBs (HMXBs), thereby indicating the great potential of this survey for revealing new XRBs.

A major challenge of finding new XRB candidates from large X-ray surveys or catalogues is contamination by X-ray emission from other objects; major contaminating X-ray emitters include cataclysmic variables (CVs), active stars/active binaries, and young stellar objects (YSOs). These objects have overlapping features with XRBs, and sometimes can only be identified via dedicated follow-up observations. A CV consists of a white dwarf accreting from a low-mass cool star, so they have accretion-induced X-rays as well as soft X-rays from the very hot white dwarf surface. X-ray luminosities are around $10^{29-31}\,\ergs$ in quiescence \citep[e.g.,][]{Reis13} but this can reach $\sim 10^{33-34}\,\ergs$ in outburst \citep{Baskill05}. Non-thermal processes in stellar coronae can emit X-rays \citep[see e.g.,][for a review]{Gudel04} with luminosities ranging between $\approx 10^{27-31}\,\ergs$ \citep[e.g.,][]{Wang20}. Coronal activity can be greatly enhanced in active binaries where the stars are tidally locked and forced to rotate rapidly; for example, RS CVn systems \citep[e.g.,][]{Walter78b, Walter78c} --- close binaries composed of a subgiant or giant star and a main-sequence or subgiant companion --- are strong X-ray sources as are close binaries of later-type dwarfs \citep[e.g., BY Dra;][]{Dempsey97}. X-ray emission has also been observed in subclasses of YSOs, typically in later-stage YSOs (e.g., T Tauri stars), which is attributed to enhanced coronal activity and accretion shocks \citep[see e.g.,][for a review]{Feigelson99}.

Without dedicated spectroscopic or time-domain observations, one effective way to distinguish between stellar coronal X-ray emission and accretion onto compact objects is from the X-ray-to-bolometric ratio ($L_\mathrm{X}/L_\mathrm{bol}$). This ratio in active stars/binaries, especially those with late-type dwarfs, does not exceed the saturation limit of $\sim 10^{-3}$ \citep{Vilhu83, Vilhu84, Vilhu87, Fleming89}, which also holds for most RS CVn systems \citep[e.g.,][]{Walter81}. Moreover, YSOs have $L_\mathrm{X}/L_\mathrm{bol}$ close to the saturation limit, between $10^{-4}$ and $10^{-3}$ \citep{Vilhu84, Vilhu87, Wright11}. In accreting compact objects, X-ray emission accounts for a much higher proportion of the system's energetics, and $L_\mathrm{X}/L_\mathrm{bol}$ is generally above the saturation limit \citep[e.g.,][]{Bernardini16}. This applies to both Galactic XRBs and to active galactic nuclei (AGNs). In actual applications, bolometric flux/luminosities is often substituted by more observable values in optical bands. For example, \citet{Tranin22} present X-ray/$r$-band flux ratios for the common X-ray emitters, clearly separating stars from XRBs and CVs. A more recent and comprehensive study is presented by \citet[][R24 hereafter]{Rodriguez24a}, who uses the \gaia\ G-band flux ($F_\mathrm{G}$) for the X-ray/optical ratios and developed an empirical relation between $\fxfg$ and the \gaia\ $\bprp$ colour to separate active stars/binaries and YSOs from accreting compact objects (more details in Sect \ref{sec:candidates_selection}). \citet{Wang25} apply a transformed version of this empirical relation (see their eq. 6) to the Data Preview 1 release of the {\it Legacy Survey of Space and Time (LSST)} from the {\it Vera C. Rubin Observatory} to search for potential COBs. Future {\it LSST} releases will offer much broader and deeper coverage for such searches.

X-ray/optical ratios have proved effective in selecting accreting compact objects, but less so in BH or NS XRB searches due to the difficulty in distinguishing them from CVs. It is, however, possible to make use of the evolutionary difference between CVs and BH or NS binaries to further refine candidate samples. The key difference is that BHs and NSs have experienced energetic supernovae, while white dwarfs in CVs have not. NKs associated with the supernovae can greatly modify the kinematics of the compact object and/or the COB as a whole. In certain instances, the parent binary can survive the supernova and be greatly accelerated \citep{Brandt95, Nelemans99}, and this imprint of NKs can be observed as high or even runaway space velocities. In fact, a number of studies have found high peculiar velocities ($\vpec$) in both isolated pulsars \citep[e.g.,][]{Lyne94, Hobbs05}, and COBs, including XRBs \citep[e.g.,][]{Mirabel01, Gandhi19, Atri19, Fortin22, ODoherty23, Zhao23, DashwoodBrown24} and BPSRs \citep[e.g.,][]{Jennings18, ODoherty23, Zhao23}. $\vpec$ is the space velocity relative to the bulk motion of the parent population, which has been broadly used as an observational proxy for NK strength.

In this work, we conduct a comprehensive search for high-velocity X-ray sources (HVXSs) using archival X-ray source catalogues and astrometric information from the \gaia\ survey. By cross-matching X-ray catalogues with \gaia, a sample of HVXSs that have high X-ray/optical ratios are selected. In Sect \ref{sec:methodology}, we introduce the X-ray source catalogues employed, cross-matching techniques, and the selection processes used to identify the HVXS sample; we also further curate a sample of gold sources that have well-constrained astrometry and robust X-ray/\gaia\ association. In Sect \ref{sec:results}, we overview properties of the selected HVXS sample. In Sect \ref{sec:discussion}, we compare the HVXSs to a control sample and discuss individual gold sources. Finally, in Sect \ref{sec:conclusions}, we summarise our findings and draw our main conclusions.



\section{Methodology}
\label{sec:methodology}
\subsection{X-ray catalogues}
The following sections summarise the construction of the final sample, which integrates data from various up-to-date X-ray source catalogues, including the \chandra\ Source catalogue \citep[\cscalias, ver. 2.1;][]{Evans24}, the \xmm\ Serendipitous Source catalogue \citep[\xmmalias, ver. DR14;][]{Webb20}, the \swift -XRT Point Source catalogue \citep[\swiftalias;][]{Evans19}, and the \erass\ source catalogue \citep[ver. DE DR1;][]{Merloni24}. This results in a raw total exceeding 2.1 million X-ray sources.

\subsection{Selection of point sources}
\label{sec:selection-confident-point-sources}
We select point sources that are confidently detected in each catalogue. For \cscalias, this corresponds to {\tt significance} $\geq 5.0$, {\tt extent\_flag=FALSE}, and {\tt conf\_flag=FALSE}. For 4XMM DR14, we set a $5\,\sigma$ detection limit ({\tt SC\_DET\_ML $\geq$ 14}; \citealt{Webb20}), only keeping those with {\tt SC\_SUM\_FLAG=0} and {\tt CONFUSED=``f"} to avoid spurious and confused sources; extended sources are excluded by constraining {\tt SC\_EXTENT $\leq$ 1\arcsec}. To clean the 2SXPS catalogue, we specify {\tt DetFlag=0} and {\tt FieldFlag=0}; this keeps confident point sources whose fields were not affected by stray light, diffuse emission, or artifacts. Finally, for the eRASS catalogue, we keep sources with {\tt DET\_LIKE} above 10, which reduces the spurious detection rate to $\approx 1\%$ \citep{Seppi22}; Extended sources are excluded by specifying {\tt EXT\_LIKE=0}. Without removing duplication, this results in a total of $\totalconfidentnumber$ confident sources.

\subsection{Calibrate X-ray positional uncertainties and removal of duplicated X-ray sources}
\label{sec:de-duplication-of-xray-catalogues}
All X-ray positional uncertainties from different catalogues are converted to a common scale corresponding to one Mahalanobis radius --- the circular radius enclosing 39.3\% of the probability in a 2D Gaussian positional error distribution. This calibrated error radius is denoted by $\rerrx$.

For the \cscalias, the semi-major ({\tt err\_ellipse\_r0}) and semi-minor ({\tt err\_ellipse\_1}) axes of the positional error ellipse represent 95\% confidence intervals\footnote{\url{https://cxc.cfa.harvard.edu/csc/columns/positions.html}}. We estimate a circular radius as the geometric mean of the two axes, $\sqrt{\mathtt{err\_ellipse\_r0 \times err\_ellipse\_1}}$. Assume an isotropic 2D Gaussian, the cumulative probability distribution function is given by
\begin{equation}
    P(< r) = 1 - e^{-r^2/2},
\end{equation}
where $r$ is in units of  Mahalanobis radii. Therefore, the conversion factor from radius enclosing 95\% probability is given by $1/\sqrt{-2\ln(1-0.95)}=0.4085$.

For the \xmmalias, the positional error radius ({\tt sc\_poserr}) encloses the true position with 63\% probability\footnote{\url{http://xmmssc.irap.omp.eu/Catalogue/4XMM-DR14/Coordinates.html}}, corresponding to $\sqrt{2}$ Mahalanobis radii. We therefore apply a conversion factor of $1/\sqrt{2}=0.7071$.

For the \swiftalias, the quoted error radius ({\tt Err90}) represents 90\% confidence under a Rayleigh distribution\footnote{\url{https://www.swift.ac.uk/LSXPS/docs.php}}, equivalent to $2.146$ Mahalanobis radius, giving a conversion factor of $0.4660$.

Finally, the \erass\ positional error radii already correspond to one Mahalanobis radius \footnote{\url{https://erosita.mpe.mpg.de/dr1}}, so no conversion is required.

We then auto-correlate the $\totalconfidentnumber$ source positions to identify and remove potential duplicates. Using the {\tt KDTree} implementation in {\sc scipy}, we efficiently search for pairs of sources whose positional error circles overlap, considering them as matches. Specifically, a pair of sources are considered identical, if their separation is smaller than the summed $\rerrx$ values. All matching pairs are then grouped into connected components using a graph constructed with the {\tt NetworkX} package \citep{networkx08}. Within each group, we retain the X-ray ID with the smallest $\rerrx$. This step removes a total of \numdiscardedbydeduplication\ sources, leaving a total of $\numafterdeduplication$ sources.

\subsection{Cross-match with \gaia}
\label{sec:cross-match-with-gaia}
The deduplicated catalogue is then cross-matched against the \gaia\ DR3 catalogue \citep{GaiaCollaborationandVallenari23}. A preliminary cross-match was performed using the \nway\ package \citep{Salvato18}, which is built on a Bayesian framework that provides match probabilities. We set the maximum matching radius to $15\arcsec$, which is greater than $99\%$ of the positional uncertainties in all four catalogues. We do not limit our cross-matching with additional prior information, so our subsequent selection processes can be based on a comprehensive sample. We then only kept the matches that have {\tt p\_any}$\geq 0.9$, {\tt p\_single}$\geq 0.9$, and {\tt match\_flag=1}. {\tt p\_any} is the probability that an X-ray source has any \gaia\ counterpart, {\tt p\_single} is the probability that compares this \gaia\ association vs. no association \citep{Budavari08}. We also {\tt match\_flag=1}, which keeps only the most confident \gaia\ counterpart for each X-ray source. After this step, we obtained a total of $\totalgaiacounterpartnumber$ confident matches.

\begin{figure}
    \centering
    \includegraphics[width=\columnwidth]{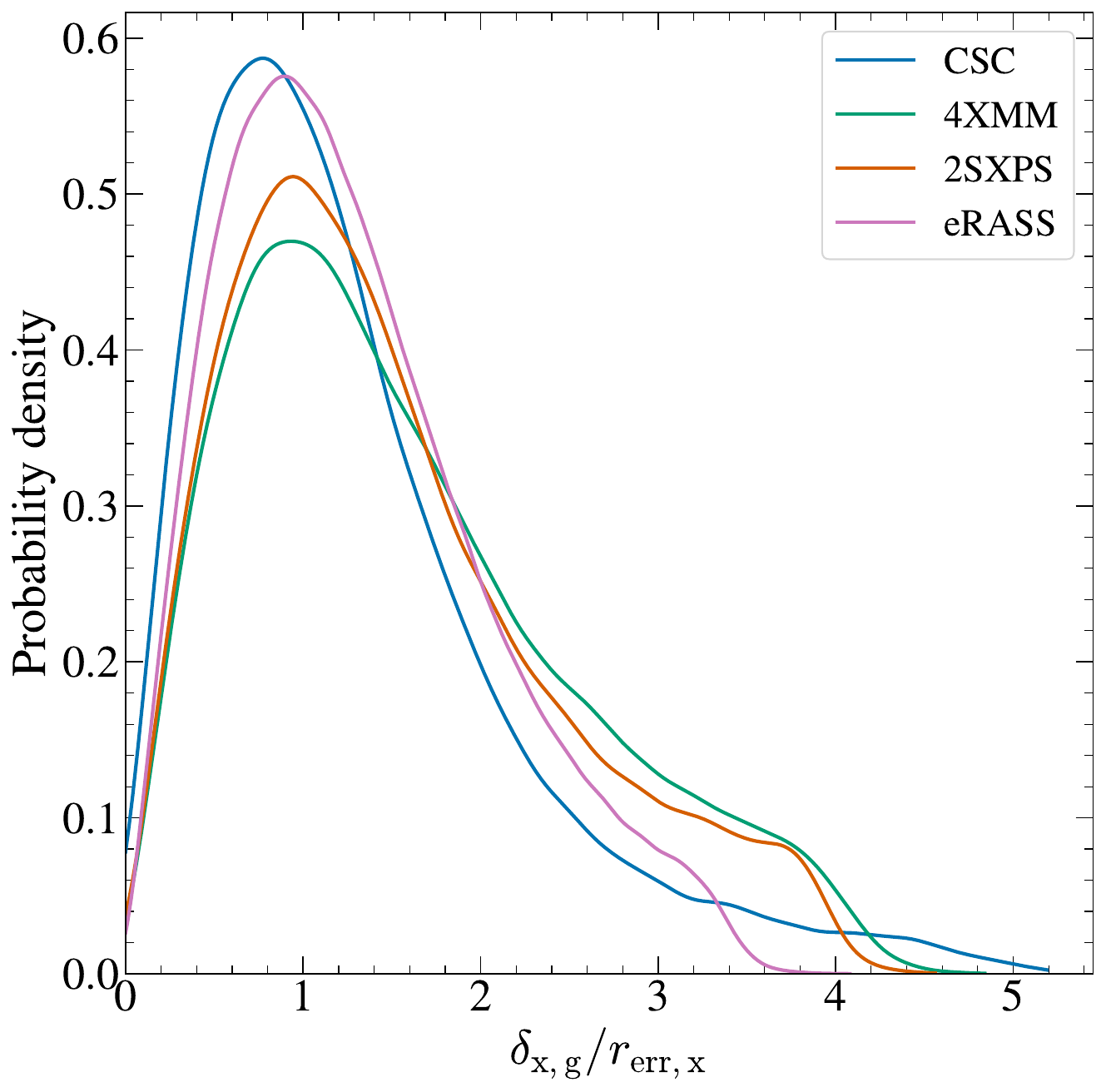}
    \caption{Distributions of \gaia/X-ray source separation (in units of positional uncertainty, i.e., $\gaiaxsep/\rerrx$) for unique one-to-one {\sc nway} matches (by X-ray source catalogues) following the cleaning process in Sect \ref{sec:cross-match-with-gaia}. The probability density is estimated by a Gaussian kernel.}
    \label{fig:sep_x_g}
\end{figure}

After the above steps, we found a total of $\totaluniqueduplicatednwaymatches$ \emph{unique} \gaia\ sources that are matched to multiple \emph{distinct} X-ray sources (Sect \ref{sec:de-duplication-of-xray-catalogues}). In fact, some high-probability {\sc nway} \gaia\ counterparts lie beyond $\rerrx$ from the X-ray positions, making one-to-many \gaia/X-ray matches possible even after the deduplication in Sect \ref{sec:de-duplication-of-xray-catalogues} and the above probability cuts. We retain only the closest pairs in terms of $\gaiaxsep/\rerrx$, removing a total of $\totalremovedduplicatednwaymatches$ matches. The cleaned \gaia/X-ray cross-matched catalogue contains $\totalgaiamatchesafterdeduplication$ one-to-one matches. Figure~\ref{fig:sep_x_g} shows the distribution of their \gaia–X-ray angular separations ($\gaiaxsep$) in units of $\rerrx$.

Two additional steps are applied to retain only the most robust matches: (1) X-ray sources with $\rerrx \ge 10\,\arcsec$ are removed, and (2) only X-ray sources whose \gaia\ counterpart is the sole match within $2\,\rerrx$ are kept. This keeps \totalgaiamatchesafterfurthercleaning\ sources.

\subsection{Producing a kinematics-worthy sample}
\label{sec:getting-a-kinematics-worthy-sample}
With \gaia\ IDs, we further clean the sample by keeping sources that have at least these astrometric parameters: coordinates ($\ra$, $\dec$), parallax ($\parallax$), and two proper motion (PM) components ($\pmracosdec$, $\pmdec{}$). These are needed for deriving their kinematic properties. Note that radial velocity (RV) is not necessary in our selection, as we will compute their space velocities in a RV-agnostic way (Sect \ref{sec:minimum-peculiar-velocity}). We use the \gaia\ ${\tt astrometric\_params\_solved}$ flag --- a 7-bit indicator of which astrometric parameters were estimated --- to select sources with valid astrometric solutions, retaining only those with ${\tt astrometric\_params\_solved} =31$, 63, or 95. 

We also removed quasi-stellar objects (QSOs) and galaxy candidates identified by the \gaia\ Discrete Source Classifier \citep[DSC;][]{Delchambre23}. The DSC provides combined class probabilities based on \gaia\ astrometry, photometry, and BP/RP spectra; for example, {\tt classprob\_dsc\_combmod\_quasar} gives the combined probability that a source is a QSO. We keep sources with an overall combined (“Combmod”) probability of being a star ({\tt classprob\_dsc\_combmod\_star}) greater than 0.9, as well as those not classified by the DSC (i.e. with no {\tt classprob\_dsc\_combmod\_star} available). This step produces a sample of \totalgaiasourceswvalidastrometryandnotagns\ \gaia\ sources.


\subsection{Distances}
\label{sec:distances}
We use the inferred distances from the \citep[][hereafter B21]{Bailer-Jones21} catalogue, prioritising the values inferred by incorporating \gaia\ parallaxes and photometry, which are called ``photogeometric" distances ($\distance$). This estimate gives generally more constrained distributions compared to those of the purely geometrically-inferred values (B21). 

The \gaia\ archive only provides point estimates (medians) and credible interval bounds, and in many cases, the upper and lower errors are asymmetric, indicating a skewed posterior. For our Monte Carlo simulation, we employ the gamma distribution to approximate the posterior, whose density distribution function is given by

\begin{equation}
    f(d) = \frac{1}{\Gamma(\alpha)\theta^\alpha}d^{\alpha-1}\exp\left(-\frac{d}{\theta}\right).
    \label{eq:gamma-distribution}
\end{equation}
Here, $\alpha$ and $\theta$ determine the shape of the distribution, and we adjust their values to fit the lower and upper bounds, while freezing the mode of the distribution to the reported median. The best fit is found by minimizing summed quadratic difference between nominal and analytical values of the bounds. 

\subsection{Minimum peculiar velocity}
\label{sec:minimum-peculiar-velocity}
We then compute $\vpec$ for the kinematics-worthy sample. In the context of this work, we assume that these sources were born in the Galactic disc, so by definition, $\vpec$ is the 3D space velocity relative to disc rotation at the source's projected radial offset from the Galactic centre. Computation is set up using the rotation curve defined under the {\tt MWPotential2014} from the {\sc galpy} package \citep{Bovy15}, adopting $\rsungc=8.34\pm 0.16\,\kpc$, the rotation speed of the local standard of rest (LSR) $\rotlsr=240\pm 8\,\kms$, and the Cartesian components of solar motion relative to the LSR, $(\usol, \vsol, \wsol) = (10.7\pm 1.8, 15.6\pm 6.8, 8.9\pm 0.9)\,\kms$ \citep{Reid14a}. 

\begin{figure}
    \centering
    \includegraphics[width=\columnwidth]{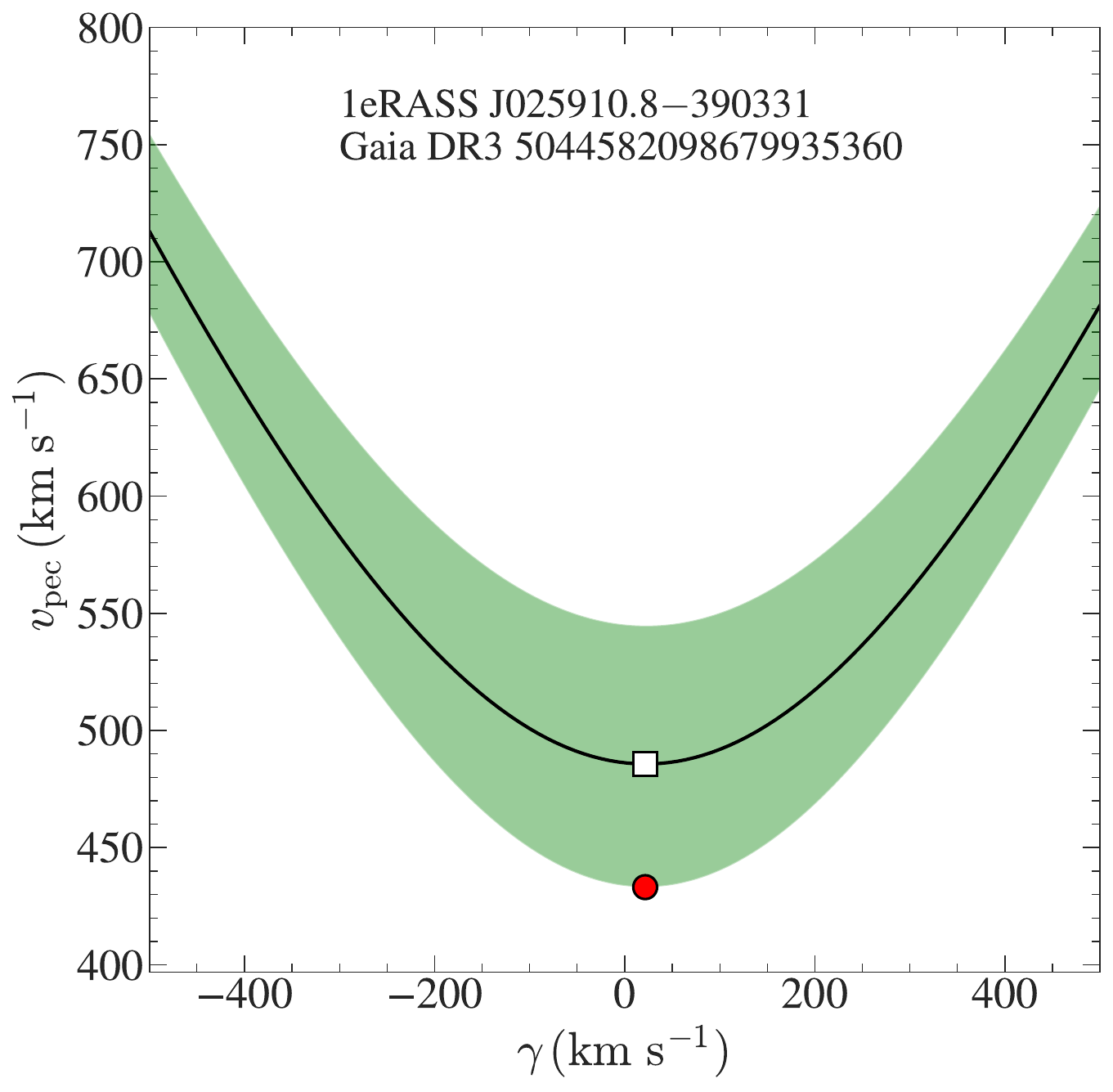}
    \caption{Peculiar velocity ($\vpec$) values at a range of systemic radial velocities ($\gamma$) for the X-ray source 1eRASS J025910.8$-$390331. The shaded area represents the $1\,\sigma$ confidence region. The minimum $\vpec$ ($\vpecmin$) and the $1\,\sigma$ lower limit of $\vpecmin$ ($\vpecminlo$) are marked by a open square and a filled red circle, respectively.}
    \label{fig:vpec_vs_gamma}
\end{figure}

We then followed the formulation in \citet{Reid09} to convert the astrometric parameters to $\vpec$. For most of our sources, there is no RV information; even though \gaia\ provides RVs for some (generally bright) sources, they do not necessarily track systemic motion of the source (in cases where the sources are not single). For non-single sources, it is the systemic RV ($\gamma$) that is needed for calculating $\vpec$. However, measuring $\gamma$ usually requires multi-epoch RV values, which is challenging to obtain on a large scale. We instead stayed agnostic to the exact $\gamma$ but assumed a broad range of values and searched for the minimum possible $\vpec$ value, denoted as the minimum peculiar velocity ($\vpecmin$), and the corresponding $\gamma$ value is denoted as $\vpecgammamin$. Uncertainties on $\vpecmin$ were propagated from all input astrometric parameters and Galactic constants following a Monte Carlo resampling method. Specifically, we drew $1000$ samples assuming a normal distribution centred on the nominal values and spread as their $1\,\sigma$ uncertainties for the proper motion components, while sampling from the approximated posterior (Sect \ref{sec:distances}). Then, from the resulting $\vpecmin$ values, we found the $16$th and $84$th percentiles as the ($\approx 1\,\sigma$) lower and upper limits. In Fig.~\ref{fig:vpec_vs_gamma}, we show an example of $\vpec$ as a convex function of $\gamma$, and the location of $\vpecmin$ and its $1\,\sigma$ lower limit ($\vpecminlo$).

\subsection{Where to draw the line?}
\label{sec:where-to-draw-the-line}
COBs are generally expected to be accelerated by NKs and move at faster $\vpec$ compared to objects that have never experienced supernovae, but how distinct are they compared to the major contaminants? To make a comparison, we use the \gaia\ astrometry of known COBs compiled in \citet{Zhao23}, including XRBs, BPSRs, and NICOBs; we also expanded this compilation by adding new NICOBs discovered by \citet{El-Badry24}. We then collate major contaminating sources from different references, including active stars \citep{Wright11}, active binaries \citep{Eker08}, confident YSO candidates \citep{Marton19}, and CVs \citep{Ritter03}. Similar to our selection process above (Sect \ref{sec:getting-a-kinematics-worthy-sample}), these catalogues are also checked against \gaia\ DR3, keeping those that have at least $5$-parameter astrometry. To then compute their $\vpec$ values, we use \gaia\ radial velocities for active stars and YSOs, systemic radial velocities compiled by \citet{Ak15} for CVs, and the centre-of-mass radial velocities from the \citet{Eker08} catalogue for active binaries. 

\begin{figure}
    \centering
    \includegraphics[width=\linewidth]{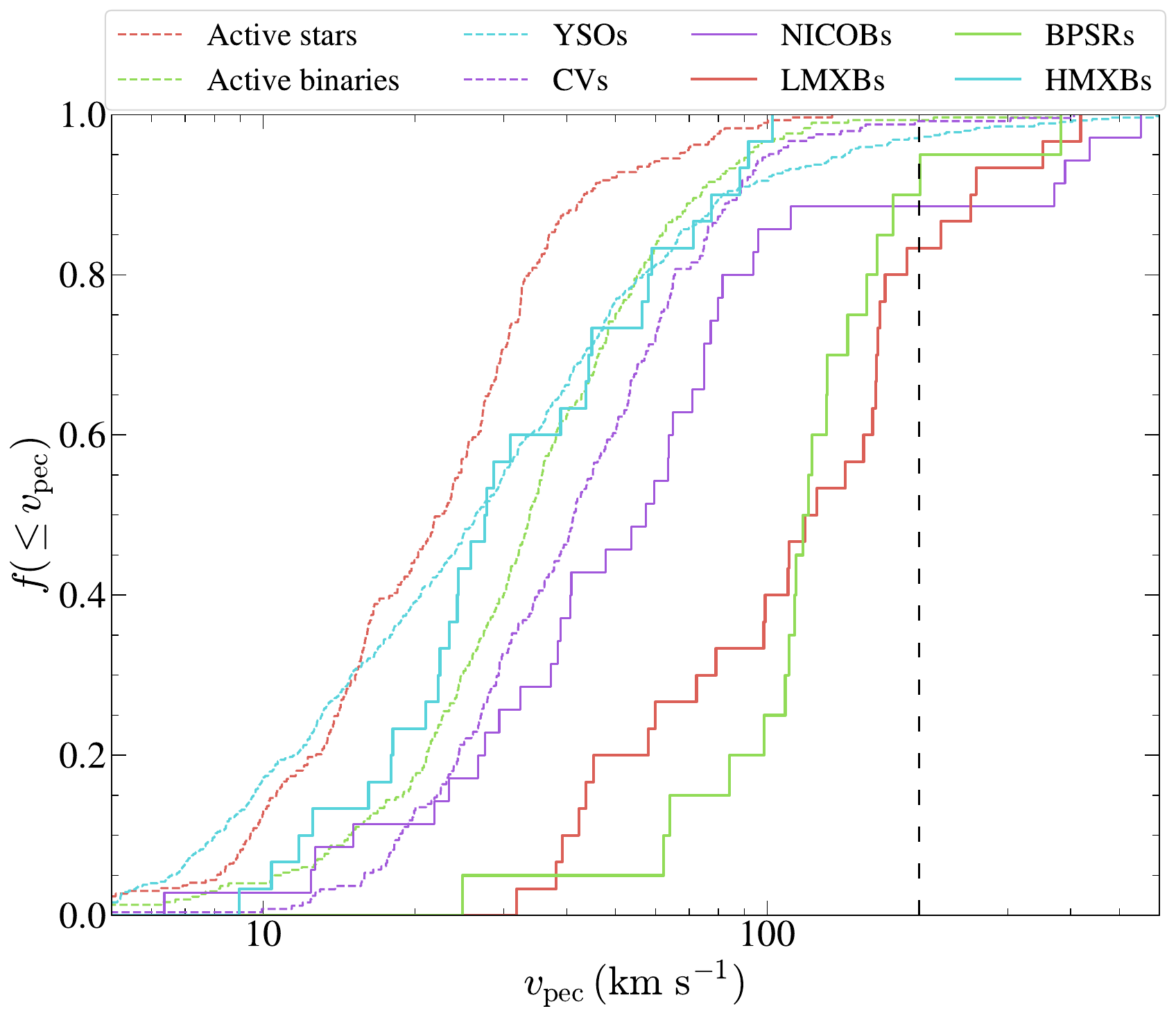}
    \caption{Empirical cumulative distribution functions of peculiar velocity ($\vpec$) plotted for types of compact object binaries, including low-mass X-ray binaries (LMXBs), high-mass X-ray binaries (HMXBs), binary pulsars (BPSRs), and non-interacting compact object binaries (NICOBs). Major contaminating sources to XRB searches are represented by dashed lines, including cataclysmic variables (CVs), active stars/binaries, and young stellar objects (YSOs). The vertical dashed line marks the lower limit of $\vpecminlim\,\kms$ we used for selecting high-velocity sources. LMXBs and BPSRs have apparently higher $\vpec$ values compared to the major contaminants.}
    \label{fig:ecdf-vpec-xrbs-and-contaminants}
\end{figure}

Fig.~\ref{fig:ecdf-vpec-xrbs-and-contaminants} compares the empirical cumulative distribution functions of $\vpec$ for COBs (XRBs, BPSRs, and NICOBs) with those for different classes of contaminating sources. While HMXBs and some NICOBs are somewhat similar to the contaminating sources, most low-mass XRBs (LMXBs) and BPSRs move at apparently higher velocities (i.e., above $\approx 100\,\kms$). We chose $\vpecminlim\,\kms$ as the limit for selecting HVXSs in this study. This limit effectively excludes over $97\%$ contaminating sources of all kinds, while maintaining sensible fractions of LMXBs ($\approx 15\%$), BPSRs ($\approx 5\%$), and NICOBs ($\approx 10\%$) (Figure \ref{fig:ecdf-vpec-xrbs-and-contaminants}).

\subsection{The HVXS sample}
\label{sec:candidates_selection}
For a more conservative selection of HVXSs, we use the $1\,\sigma$ lower limit on $\vpecmin$ ($\vpecminlo$), selecting sources with $\vpecminlo\geq \vpecminlim\,\kms$ and keep sources that have a $1\,\sigma$ lower limit of X-ray/\gaia\ G-band flux ratios ($\fxfg$) above the empirical separatrix in R24. This gave a sample of $\numhvxbeforesimbad$ sources. Additional cleaning was performed on this sample in order to obtain higher-confidence COB candidates.
\begin{enumerate}
    \item The first step is to remove known COBs (either identified as XRBs or PBSPs) and non-COBs (e.g., AGNs, CVs, YSOs, CVs, etc.). To do this, we cross-matched the sample against the SIMBAD database with a search radius of $10\,\arcsec$. This yielded a total of $\numhvxsimbadmatches$ matches. Extragalactic contaminants and their candidates (e.g., {\tt Galaxy}, {\tt QSO}, {\tt AGN}, etc.) account for $\approx 37\%$ of this sample, while Galactic ones (e.g., {\tt YSO}, {\tt CataclyV*}, {\tt BYDraV}, {\tt RSCVnV*}, etc.) account for $\approx 10\%$. The crossmatch reveals a minor fraction ($\approx 7\%$) of known COBs and their candidates, consistent with their general rarity. Stars ({\tt Star}), the most common classifications ($\approx 12\%$), were kept. In addition, some classification --- such as generic X-ray sources ({\tt X} $\approx 8\%$), and high-proper-motion stars ({\tt HighPM*}; makes $\approx 2\%$) --- are also retained, as they directly align with our selection of HVXSs. The rest of the matches fall into ambiguous types that could plausibly host NSs or BHs. These are kept in our sample and summarised in Table \ref{tab:kept-simbad-types}. After this step, we are left with $\numhvxaftersimbad$ sources.

    \item Our method of computing $\vpec$ assumes a Galactic disc origin, which does not apply to sources associated with the halo. While it is challenging to systematically remove halo sources, some of the sources have been astrometrically identified as cluster members. To eliminate cluster members, we perform a preliminary filtering by removing sources with indicative SIMBAD types and names (e.g., {\tt Cl, ClG}, etc.). We further compared the sample \gaia\ {\tt source\_id} with those of likely open/globular cluster members in the \citet{Hunt23} catalogue. This removes $\numclustermembers$ likely cluster members from our sample, leaving us with $\numhvxafterclustercleaning$ sources.

    \item We also removed likely members of the Small and the Large Magellanic Cloud (SMC and LMC). To do so, we cross-matched HVXS \gaia\ {\tt source\_id}s with the SMC and LMC membership catalogues from \citet{JimenezArranz23a, JimenezArranz23b}. These studies provide membership probabilities for sources near the galaxies, and the distributions shows a clear bimodal pattern: one peak near 1 (high-confidence members) and another near zero (non-members). To ensure a conservative selection, we adopted a low probability threshold (0.01 for SMC and 0.002 for LMC) to remove likely members, effectively including most of the non-member peak near zero. This step gives a catalogue of \numhvxaftersmclmccleaning sources.


    \item In a final step, we remove astrometric solutions that are likely spurious using the fidelity parameter derived by \citet{Rybizki22}. This is derived from a neural network classifier trained on a broad set of \gaia\ parameters, and it takes a value between 0 and 1 to quantify the probability that a \gaia\ astrometric solution is good. Following their recommendation for optimal completeness, we use the updated parameter ({\tt fidelity\_v2}) and chose $0.5$ as the threshold \citep[see figure a17 of][]{Rybizki22}. This step reduces the sample size to \numhvxaftercleaning.
\end{enumerate}
These selection steps toward the final HVXS sample are summarised in a flowchart in Fig.~\ref{fig:flow-chart-hvxs}.

\subsection{HVXS quality flags}
\label{sec:hvxs_quality_flags}
We develop a 3-digit bitmask to encode the combinations of quality criteria ({\tt quality}) satisfied by each HVXS. The design of these criteria is motivated by minimising confusion in positional matches, while ensuring a confident distance estimate which $\vpecmin$ is strongly dependent on. From the most (leftmost) to the least significant (rightmost) bit, the corresponding conditions are
\begin{itemize}
    \item $\gaiaxsep \leq \rerrx$: \gaia\ counterpart within the $1\,\sigma$ X-ray error circle
    \item $\parallax/\sigma_\parallax\geq 5$: $\parallax$ uncertainty less than $20\%$
    \item $|\dinv - \distance| / \distance \leq 0.2$: $\distance$ within $20\%$ of $\dinv$
\end{itemize}
where $\dinv=1/\parallax$ is the distance estimate by simply inverting the zeropoint-corrected $\parallax$. Note that meeting the second criterion will automatically exclude sources with negative $\parallax$ values. The third criterion alone is not a strong quality flag but is a necessary check when $\parallax$ is relatively well-constrained (i.e., when the second and third conditions are both met). Combinations of criteria and their quality flags are listed in Table \ref{tab:quality-flags}.

\begin{table}
    \centering
    \caption{Combinations of criteria and their quality flags (Sect \ref{sec:hvxs_quality_flags}).}
    \resizebox{\columnwidth}{!}{
    \begin{tabular}{c|c|c|c|l}
    \hline
     &  $\gaiaxsep\leq \rerrx$ & $\parallax/\sigma_\parallax\geq 5$ & $\frac{|d_\mathrm{inv} - \distance|}{\distance} \leq 0.2$ & $N$\\
    \hline
    $000_2=0$ & & & & 1450 \\
    \hline
    $001_2=1$ & & & \checkmark & 73\\ 
    \hline
    $010_2=2$ & & \checkmark & & 6 \\
    \hline
    $011_2=3$ & & \checkmark & \checkmark & 20\\ 
    \hline
    $100_2=4$ & \checkmark & & & 778 \\
    \hline
    $101_2=5$ & \checkmark & & \checkmark & 30 \\
    \hline
    $110_2=6$ & \checkmark & \checkmark &  & 5\\ 
    \hline
    $111_2=7$ & \checkmark & \checkmark & \checkmark & 10 \\
    \hline
    \end{tabular}
    }
    \label{tab:quality-flags}
\end{table}


\subsection{The control sample}
\label{sec:control-sample}
We build a control sample  to explore the effect of $\vpecmin$ and $\fxfg$ selection. We start from the kinematic-worthy sample (Sect \ref{sec:getting-a-kinematics-worthy-sample}) and follow the same cleaning steps as for the construction of the HVXS sample (Sect \ref{sec:candidates_selection}). After cleaning, we remove the HVXSs to obtain the control sample with a total of $\controlsamplesize$ sources.


\subsection{A gold sample}
\label{sec:prime-sample}

We build a gold sample from the HVXSs with quality flag of 7 (i.e., all three criteria in Sec \ref{sec:hvxs_quality_flags} are met), which gives a total of $\goldsamplesizebeforecuration$ sources. The gold sources all have robust {\sc nway} matching probability {\tt p\_single} and {\tt p\_any} above $99\%$, given the \gaia\ source density. We then performed a further curation of this sample, scrutinising the field around each individual source using available images from Pan-STARRS DR1 \citep{Chambers16}, the DESI Legacy Imaging Survey \citep{Dey19}, the DECam Plane Survey \citep[DECaPS;][]{Schlafly18}, the SkyMapper Southern Survey \citep{Onken24}, and the {\it Galaxy Evolution Explorer} \citep[\galex;][]{Martin05}. This step is to further remove ambiguous matches, especially in very crowded regions where faint interlopers are within $\rerrx$ but not in \gaia.


After the above checks, we focus on a final total of \goldsamplesize\ sources for further discussion in this work. However, since the exclusion process is mostly based on the matching confidence, it does not negate the potential physical significance of the remaining sources. Notably, the gold sample contains only \erass\ sources after the curation, with a median $\rerrx$ of a $4\arcsec$. X-ray observations with precision localisation (e.g., \chandra/ACIS) would significantly reduce confusion.

\subsection{Likely non-single sources}
\label{sec:non-single-sources}
\gaia\ provides the \aengaia\ ($\aen$), an astrometric indicator of source non-singularity. This measures ``astrometric wobble", which is quantified by the deviation from a standard astrometric (5-parameter) fit. A ``well-behaved" source would have $\aen$ around zero, while a large positive $\aen$ indicates that the fit residuals deviate from the expected observational noise. Typically, $\aen$ is considered significant when its \aensiggaia\ ($\aensig$) is greater than $2$ \citep{GaiaCollaborationandVallenari23}. This deviation could be the result of binary orbital motion, so could be used as one measure of source binarity.

In our HVXS sample, about $54\%$ have positive $\aen$, and $\approx 20\%$ of these positive $\aen$ values are significant ($\aensig\geq 2$). $\aen$ can be treated as a measure of astrometric wobble and therefore an estimate of the semi-major axis of binaries. \citet{Gandhi22} used $\sqrt{2}\aen d$ as an estimate of semi-major axes assuming that the wobble is due to orbital motion. This estimate could be very different from actual values for some binaries and should only be interpreted with caution.

\begin{figure*}
    \centering
    \includegraphics[width=\textwidth]{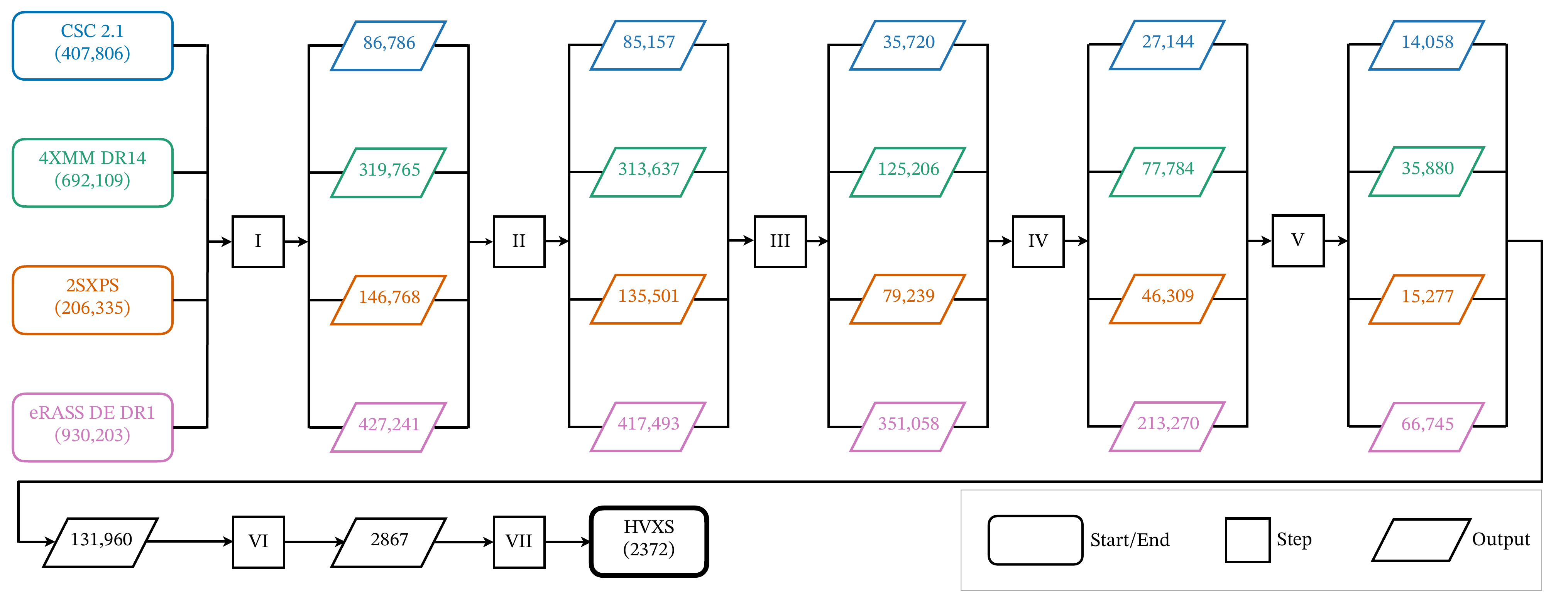}
    \caption{A flowchart demonstrating steps (represented by squares) toward the establishment of the final HVXS sample. Numbers of sources before and after each step are shown within the shapes. I: keep only confident point sources (Sect \ref{sec:selection-confident-point-sources}); II: deduplication of X-ray catalogues (Sect \ref{sec:de-duplication-of-xray-catalogues}); III: cross-match with \gaia\ using {\sc nway}, keeping only the high-probability and confident matches (Sect \ref{sec:cross-match-with-gaia}); IV: keep only the closest X-ray/\gaia\ matches (Sect 
    \ref{sec:cross-match-with-gaia}); further cleaning of the cross-matched catalogue, keeping X-ray sources whose (1) $\rerrx\leq 10\,\arcsec$, (2) \gaia\ counterpart is within $2\,\rerrx$, and (3) \gaia\ counterpart is the only \gaia\ source within $2\,\rerrx$; V: keep sources that have at least 5 astrometric parameters, removing likely QSOs and galaxies (Sect \ref{sec:getting-a-kinematics-worthy-sample}); VI: select based on $\vpecminlo\geq \vpecminlim\,\kms$ and $\fxfg$ above the R24 empirical separatrix (Sect \ref{sec:candidates_selection}); VII: further cleaning (Sect \ref{sec:candidates_selection}).}
    \label{fig:flow-chart-hvxs}
\end{figure*}

\section{Results}
\label{sec:results}
\begin{figure*}
    \centering
    \includegraphics[width=0.8\textwidth]{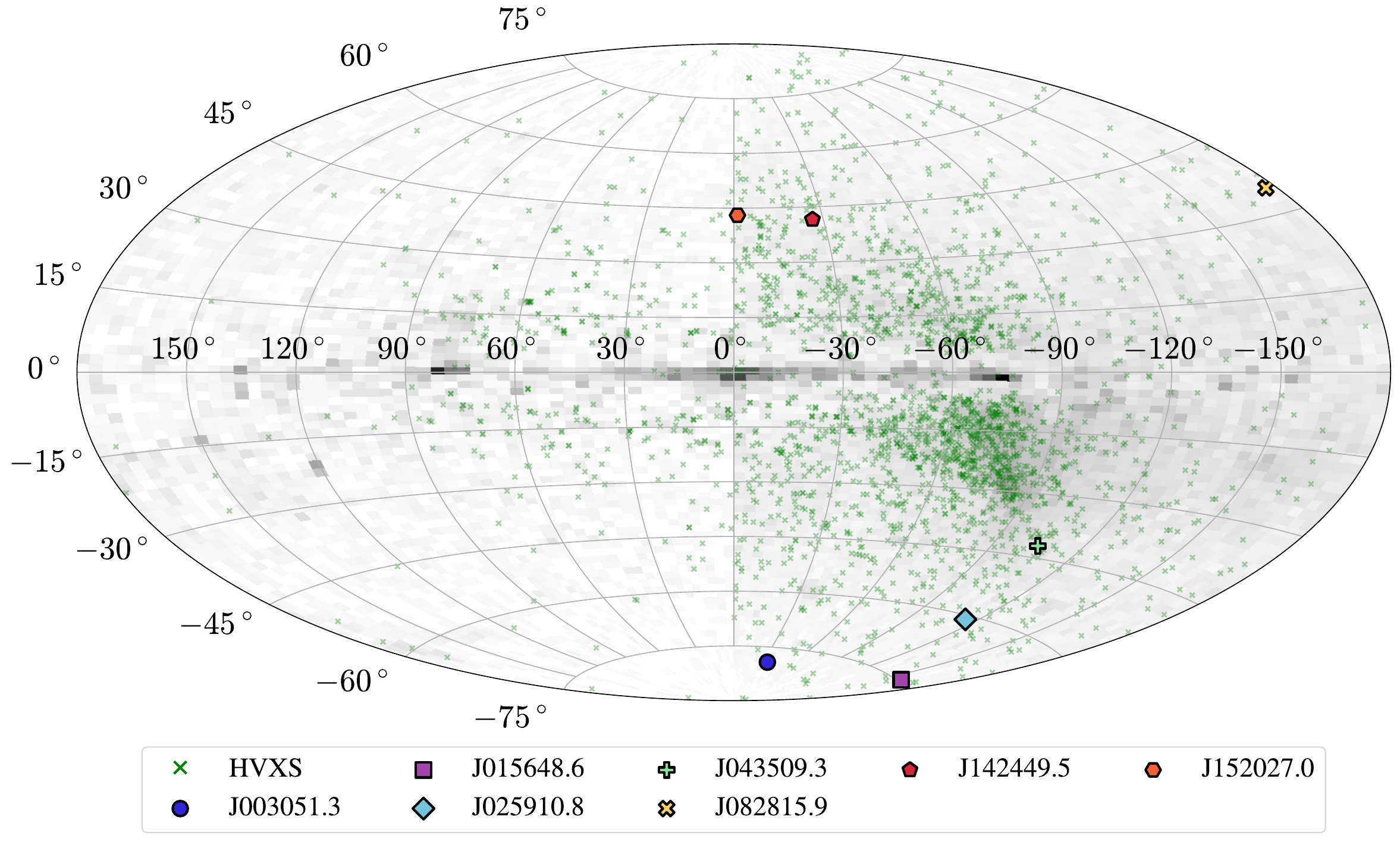}
    \caption{Galactic map showing the density distribution of the control sample (binned background in grey color scale) and positions of the selected HVXSs (green crosses). The western Galactic hemisphere ($-180^\circ \leq l\leq 0^\circ$) is more abundant of HVXSs due to the comprehensive coverage of the eRASS-DE DR1. Sources from the gold sample are indicated by markers of distinct shapes and colours.}
    \label{fig:hvxs-galactic-map}
\end{figure*}

Table \ref{tab:summary-of-samples} collates information of different samples, and key properties of the gold sources (Sect \ref{sec:prime-sample}) are listed in Table \ref{tab:gold-sources}. Sky positions of the HVXS and gold samples are presented in a Galactic map in Fig.~\ref{fig:hvxs-galactic-map}, where the number density of the control sample is binned and displayed in the background. The western Galactic hemisphere is clearly more abundant in HVXSs because of eRASS-DE DR1's comprehensive coverage.

\begin{figure*}
    \centering
    \includegraphics[width=\textwidth]{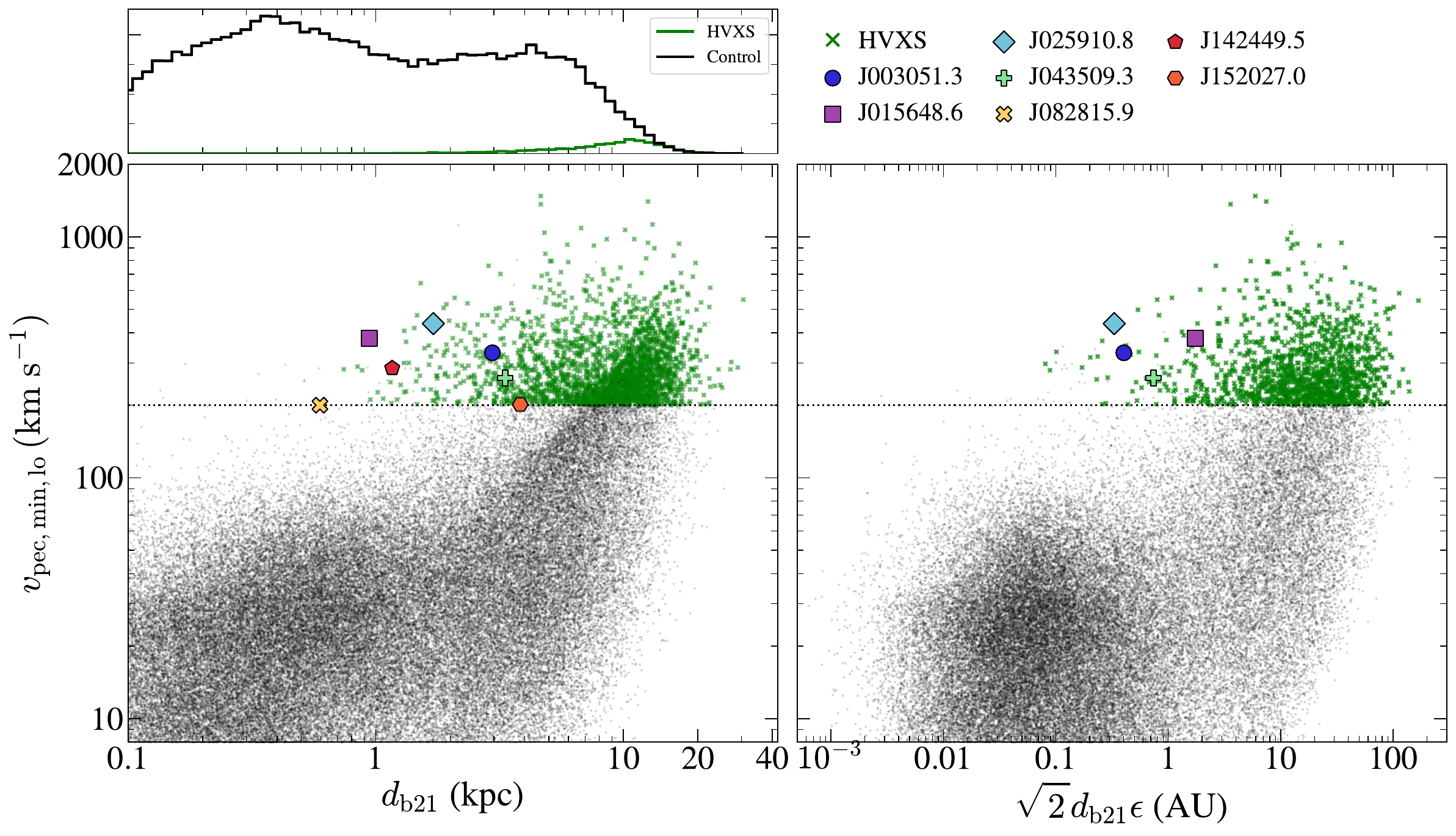}
    \caption{{\it Left}: Distribution of estimated distances (median values) and $1\,\sigma$ lower limits on $\vpecmin$ shown for the control sample (black points) and the HVXS sample (green crosses). The dotted horizontal line marks the $\vpecminlim\,\kms$ limit on $\vpecminlo$ that we used to select the high-velocity sources; the top panel exhibits the $\distance$ distributions for the HVXS and control sample. {\it Right}: The same $\vpecmin$ lower limit plotted against estimates on binary semi-major axis for sources with non-zero astrometric excess noise ($\aen$).}
    \label{fig:dist-vs-vpec-min-lolim}
\end{figure*}

Fig.~\ref{fig:dist-vs-vpec-min-lolim} displays the distribution of $\distance$ and $\vpecminlo$. The control sample exhibits a large peak around $0.4\,\kpc$ and a broad hump from $2-6\,\kpc$. Overall, there is an upward trend of $\vpecminlo$ as a function of $\distance$, so our selection on $\vpecminlo$ greatly reduces nearby ($\leq 1\,\kpc$) sources. The right panel shows distribution against $\aen$, and HVXSs spread across different regions of $\aen$ values, i.e., even relatively wide binary candidates can have large space velocities. Wide binaries are especially intriguing if they are moving at high space velocities, as their progenitors could be more prone to disruption by strong NKs.

\begin{figure}
    \centering
    \includegraphics[width=\columnwidth]{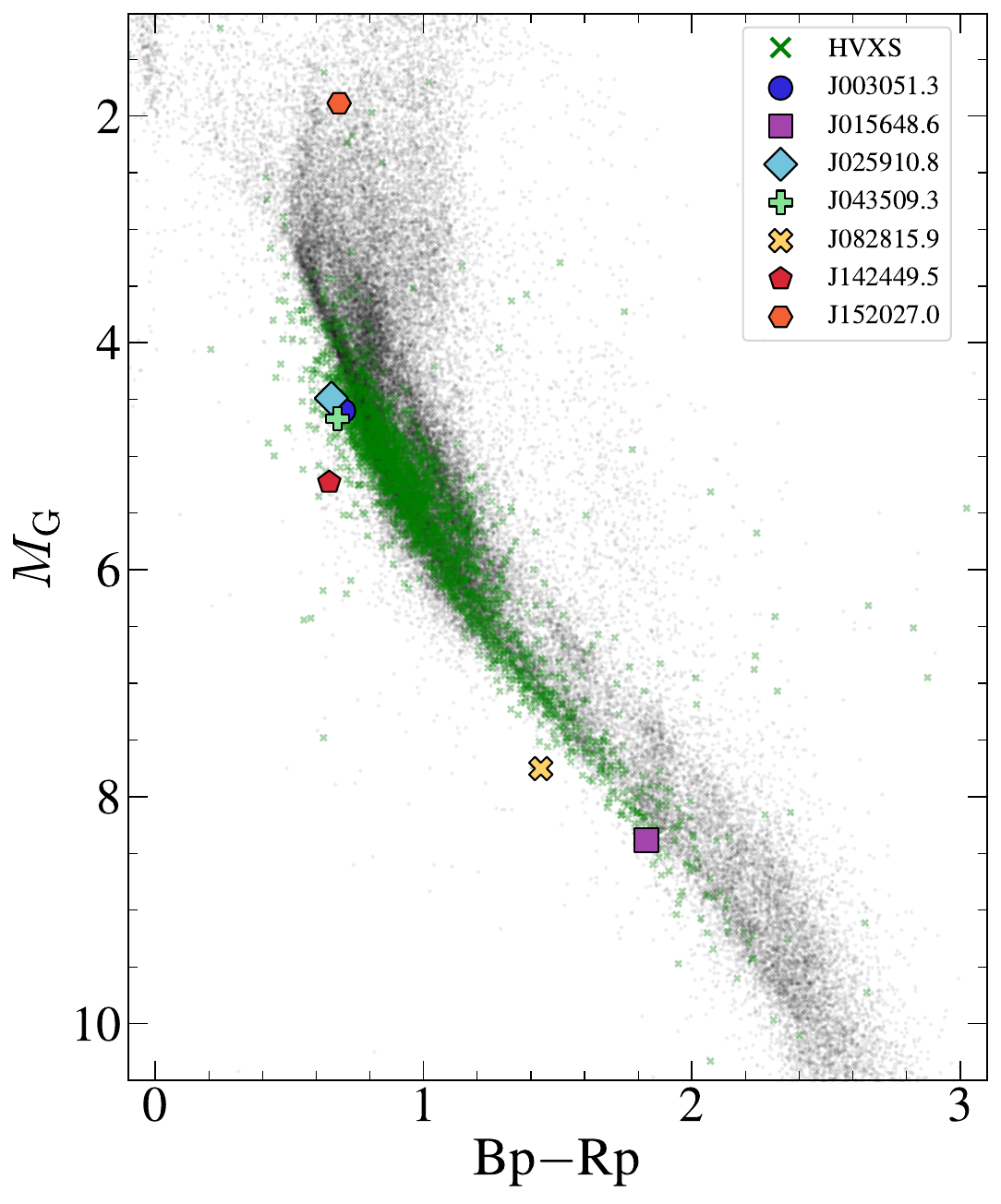}
    \caption{\gaia\ colour-magnitude diagram plotting $\bprp$ colour vs. G-band absolute magnitude ($\gabs$) for the HVXS (green crosses) and the control sample (grey points; Sect \ref{sec:control-sample}). Sources from the gold sample are indicated by markers of distinct shapes and colours. Colours and magnitudes are corrected using available reddening and extinction values from GSP-Phot.}
    \label{fig:gaia-cmd}
\end{figure}

Fig.~\ref{fig:gaia-cmd} gives a photometric overview of the samples, showing a \gaia\ $\bprp$ colour-magnitude diagram (CMD) of the control, the HVXS, and the gold samples. $\bprp$ colours and absolute G-band magnitude ($\gmag$) are only partly corrected for extinction for sources with available reddening and extinction estimates from \gaia's General Stellar Parametrizer from Photometry module \citep[GSP-Phot;][]{Andrae23}. Overall, most HVXSs are consistent with the control sample, but the latter has a broader distribution of colours. The HVXS has $\bprp$ colours between $0.6$ and $2.2$ ($95\%$ equal-tail interval), corresponding to spectral types between F6V and M2V \citep{Pecaut13}, while the control sample hosts apparently more redder M dwarfs and giants. We also note that there is a slight blue excess of the HVXSs relative to the control sample around $\gmag$ between $4$ and $5$, which is in line with our $\fxfg$ selection of sources with prominent X-ray emission. 

\begin{figure*}
    \centering
    \includegraphics[width=0.8\textwidth]{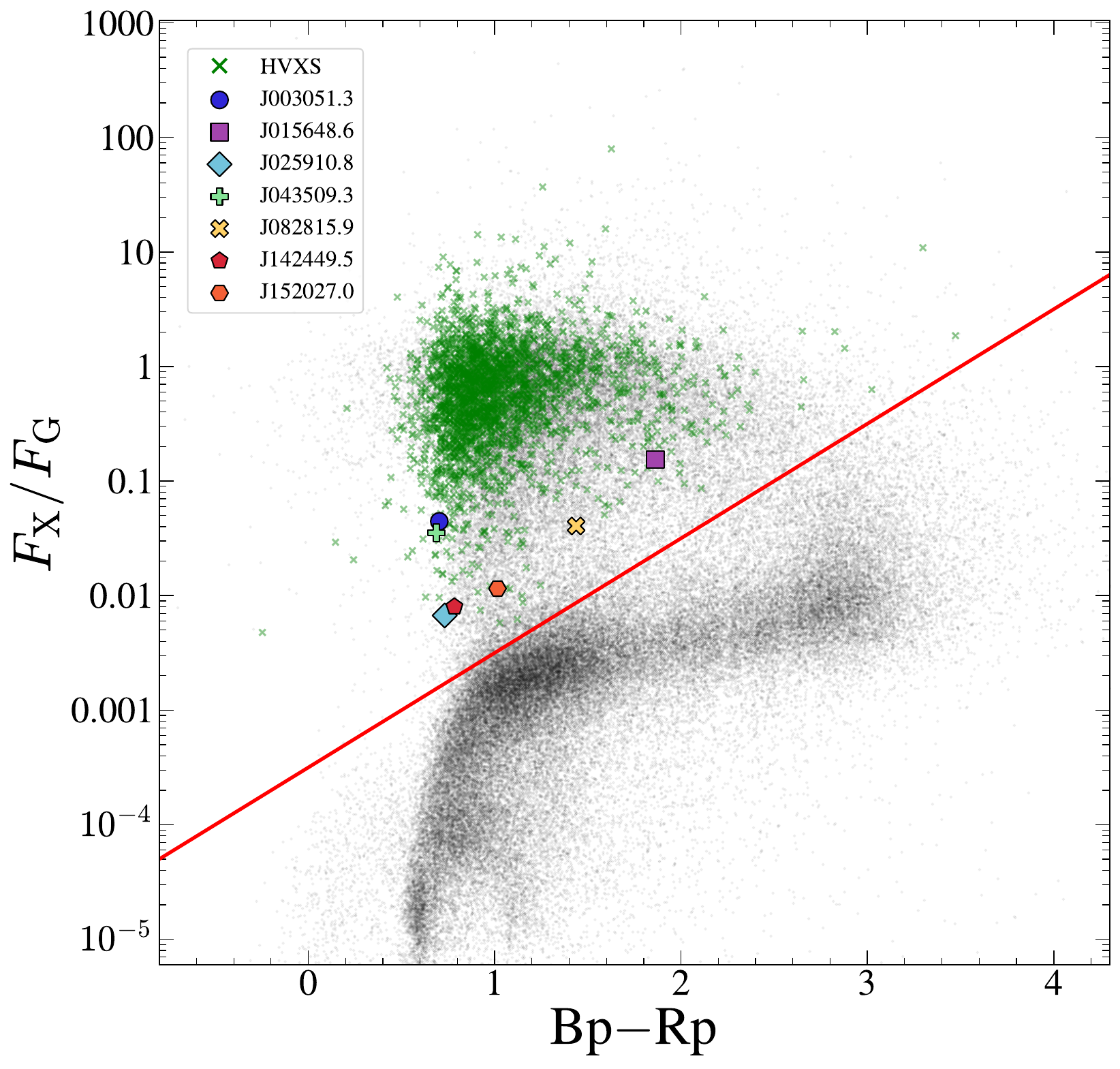}
    \caption{X-ray to \gaia\ G-band flux ratio ($\fxfg$) vs. \gaia\ $\bprp$ colours plotted for COB candidates. Black points are sources in the control sample. Sources from the gold sample are indicated by markers of distinct shapes and colours. The solid line marks the R24 empirical separatrix: $\log_{10}(\fxfg) = (\bprp)-3.5$.}
    \label{fig:fxfg_vs_bprp}
\end{figure*}

An overview of our samples' $\vpecmin$ and $\fxfg$ values are presented in Fig.~\ref{fig:fxfg_vs_bprp}. The solid line in each panel depicts the empirical relation from R24. Note that the X-ray bands are heterogeneous among different catalogues; the bands are relatively broader and similar for \cscalias\ ($\cscband$), \xmm\ ($\xmmband$), and \swiftalias\ ($\swiftband$), while we use the more sensitive soft ($\erassband$) band for \erassalias, so the $\fxfg$ of \erassalias\ sources should be treated as lower limits when compared to the other sources.

\begin{figure*}
    \includegraphics[width=\textwidth]{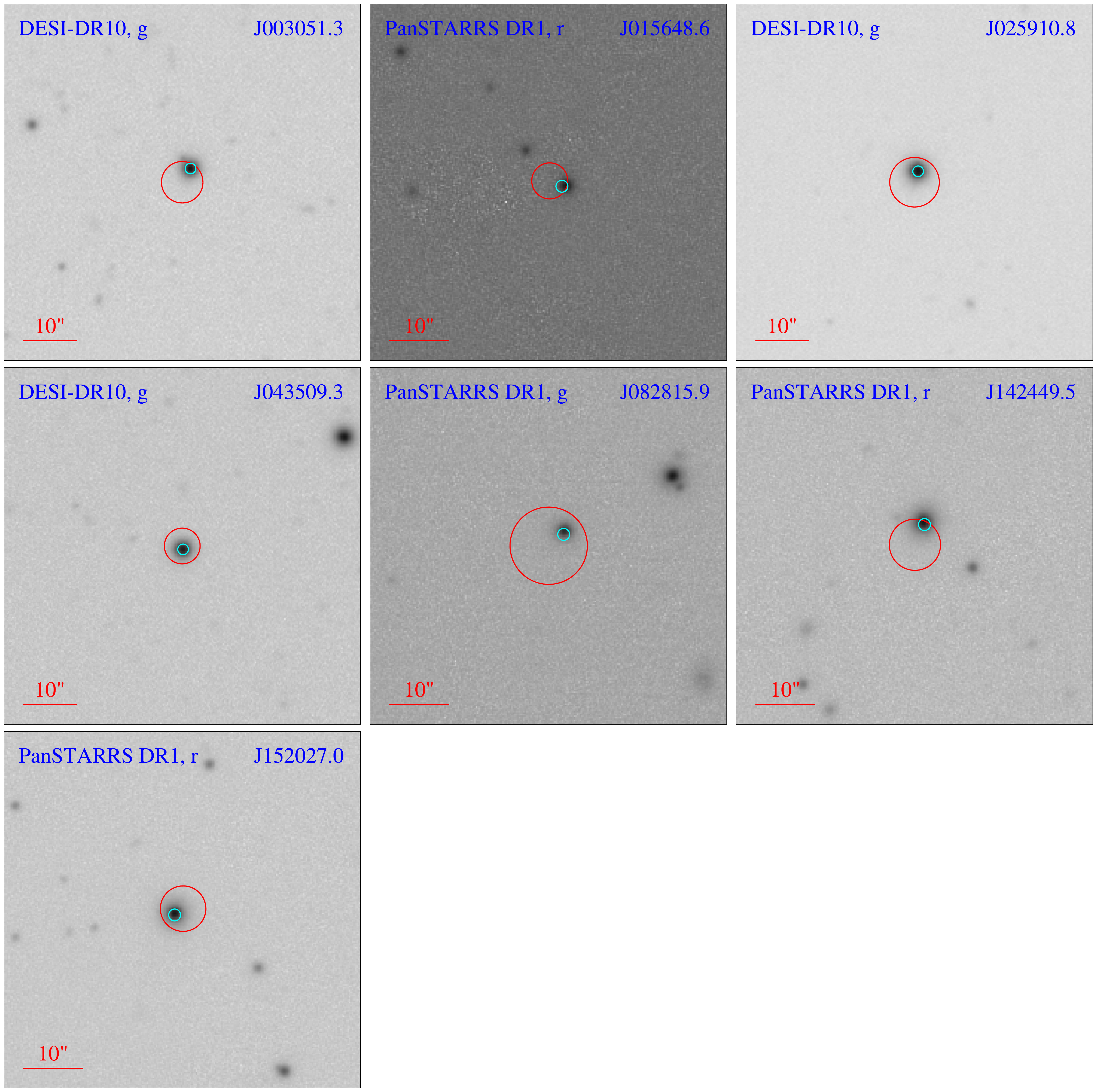}
    \caption{Optical finding charts for $\goldsamplesize$ gold sources showing $60\arcsec\times 60\arcsec$ fields centered on the X-ray positions; north is up and east is to the left. The X-ray error radius ($\rerrx$) is shown by the red circle, and the \gaia\ counterpart is indicated by a cyan circle.}
    \label{fig:gold-source-finders}
\end{figure*}

Fig. \ref{fig:gold-source-finders} presents optical finding charts for the $\goldsamplesize$ gold sample sources. These figures are referred to in the discussion of individual gold sources.

\begin{table*}
\centering
\caption{A summary of samples}
\label{tab:summary-of-samples}
    \begin{tabular}{llll}
    \hline
    Sample  & \# of sources & Description & Section \\
    \hline
    HVXS    & \numhvxaftercleaning & $\vpecminlo\geq \vpecminlim\,\kms$ and $\fxfg$ above the R24 empirical limit. & \ref{sec:candidates_selection} \\
    Control & \controlsamplesize   & Control sample chosen as a complement to the HVXS sample & \ref{sec:control-sample} \\
    Gold    & \goldsamplesize      & Curated sources from HVXS with {\tt quality}=7 & \ref{sec:prime-sample} \\
    \hline
    \end{tabular}
\end{table*}

\renewcommand{\arraystretch}{1.3}
\begin{table*}
    \caption{Summary of the $\goldsamplesize$ curated gold sources (Sect \ref{sec:prime-sample})}
    \centering
    \begin{threeparttable}
    {\small
    \begin{tabular}{c c c c c c c c c}
        \hline
        X-ray ID & \gaia\ DR3 & $\gaiaxsep$ & \gaia\ $\gmag$ & $F_\mathrm{X}$ & Sig & $\distance$ & $\vpecmin$ & $\vpecgammamin$ \\
                 &  & ($\rerrx$)  &  & ($10^{-14}\,\ergscm$) &  & ($\kpc$) & $(\kms)$ & $(\kms)$ \\
        (1) & (2) & (3) & (4) & (5) & (6) & (7) & (8) & (9) \\ 
        \hline
        1eRASS J003051.3$-$370912 & 5003862819415658112 & 0.8 & 16.96  & $6.9\pm 2.9$ & 11.1 & $3.0^{+0.5}_{-0.4}$ & $394.8^{+64.6}_{-65.1}$ & $7.1^{+0.3}_{-0.3}$ \\
        1eRASS J015648.6$-$224326 & 5134745277676212608 & 0.7 & 18.31  & $6.8\pm 2.2$ & 24.8 & $0.9^{+0.2}_{-0.2}$ & $517.7^{+156.0}_{-138.1}$ & $13.4^{+0.2}_{-0.2}$ \\
        1eRASS J025910.8$-$390331 & 5044582098679935360 & 0.5 & 15.79  & $3.0\pm 1.3$ & 10.7 & $1.7^{+0.1}_{-0.1}$ & $487.4^{+51.7}_{-51.0}$ & $22.0^{+0.6}_{-0.6}$ \\
        1eRASS J043509.3$-$481751 & 4788212717642740480 & 0.2 & 17.30  & $4.0\pm 1.0$ & 29.1 & $3.3^{+0.5}_{-0.5}$ & $292.3^{+42.7}_{-32.9}$ & $38.8^{+3.5}_{-2.7}$ \\
        1eRASS J082815.9$+$383457 & 911214901400669824  & 0.5 & 16.63  & $8.5\pm 3.6$ & 11.0 & $0.60^{+0.02}_{-0.02}$ & $238.9^{+45.9}_{-38.7}$ & $5.1^{+0.1}_{-0.1}$ \\
        1eRASS J142449.5$-$153938 & 6298627607542697088 & 0.9 & 15.81  & $3.6\pm 1.5$ & 10.8 & $1.2^{+0.1}_{-0.1}$ & $330.7^{+48.9}_{-45.7}$ & $-16.9^{+1.3}_{-1.4}$ \\
        1eRASS J152027.0$-$030609 & 4414038018672780416 & 0.5 & 15.41  & $7.4\pm 2.8$ & 12.8 & $3.8^{+0.4}_{-0.4}$ & $225.4^{+30.9}_{-23.9}$ & $-16.0^{+0.4}_{-0.4}$ \\
        \hline
    \end{tabular}
    \begin{tablenotes}
        \item (1) X-ray IAU name. (2) $\gaia$ DR3 ID. (3) X-ray-\gaia\ angular separation in units of X-ray error radii. (4) $\gaia$ $\gmag$-band magnitude. (5) X-ray fluxes between $\erassband$. (6) The $0.2-2.3\,\kev$ detection likelihood (\texttt{det\_like\_0}) of \erass. (7) Photogeometric distance estimates from B21. (8) Minimum peculiar velocity. (9) The systemic radial velocity that minimises $\vpec$.
    \end{tablenotes}
    }
    \end{threeparttable}
    \label{tab:gold-sources}
\end{table*}
               
\section{Discussion}
\label{sec:discussion}

\subsection{Individual gold sources}
\label{sec:discussion-extreme-sources}
Here we present a discussion of individual sources in the gold sample (Sect. \ref{sec:prime-sample}). For each source, we focus on the robustness of the match to its \gaia\ counterpart, the value of $\vpecmin$, photometry, and possible binarity (Sect. \ref{sec:non-single-sources}). Throughout the discussion, all reported $\parallax$ values have been corrected for zeropoint offsets. Sources are referred to using the ``RA" part of their X-ray identifier (e.g., 1eRASS J003051.3$-$370912 is referred to as J003051.3).

{\bf 1eRASS J003051.3$-$370912}'s \gaia\ counterpart is a bright, blue source. There is a hint of a very faint source extended from the northeast of the counterpart, but it is not detected in \gaia. The photogeometric $\distance$ agrees well with $\dinv$, at approximately $3.0\,\kpc$, placing the source slightly bluer than the main sequence\footnote{Here, the main sequence refers to that defined by the control sample.} in the CMD. The $\aen$ is non-zero ($0.1\,\mas$), but not significant enough to suggest non-singularity. The well-constrained $\distance$ and PM yield a relatively robust, high $\vpecminlo$ exceeding $300\,\kms$. Combined with the high $\fxfg$ ratio, this suggests a very peculiar X-ray emitting system. It is currently $\approx 1.3\,\kpc$ from the Galactic plane so could be a halo source.

{\bf 1eRASS J015648.6$-$224326}'s X-ray error circle marginally overlaps with its \gaia\ counterpart. The only other nearby optical source is a faint, blue object located about $2\rerrx$ north-east of the X-ray position, and a faint \galex\ source is also positionally coincident with this source, the UV emission can partly contribute to its blue colour. However, associating it with the X-ray source would require more precise localisation (with an instrument such as \chandra/ACIS). For now, we consider the nearest source to be the true counterpart. The B21 estimate places the counterpart at $\distance\approx 0.9\,\kpc$, in good agreement with $\dinv$. The source still has a high $\vpecminlo$ of $\approx380\,\kms$ even at this relatively close distance. Another feature that makes this counterpart particularly interesting is its significant $\aen$ value ($1.3\,\mas$), which could be translated to a semi-major axis of approximately $1\,\mathrm{AU}$ (Fig.~\ref{fig:dist-vs-vpec-min-lolim}). This estimate is wider than typical XRBs ($\lesssim 1\,\mathrm{AU}$), but should be taken cautiously as it is sensitive to the assumption that the astrometric wobble is entirely due to orbital motion. If further observations confirm the orbital parameter, it would then be particularly interesting to explore how such wide binaries could have been accelerated and still survived being disrupted either by a strong natal kick or by dynamical processes.

{\bf 1eRASS J025910.8$-$390331}'s \gaia\ counterpart is bright ($\gmag=15.8$) and well-separated from other sources in the field. Both B21's photogeometric and $\dinv$ give a $\distance$ of $\approx 1.7\,\kpc$, and this, together with the $\bprp$ colour, places it bluer than the upper main sequence (Fig.~\ref{fig:gaia-cmd}). Its apparent PM makes it the fastest-moving source (in terms of $\vpecminlo$) among the gold sample sources, with $\vpecminlo \approx 436\,\kms$, making it a strong candidate for a runaway object escaping the Galactic potential. The \gaia\ counterpart also shows a non-zero $\aen$ ($0.14\,\mas$) that is moderately significant. 


{\bf 1eRASS J043509.3$-$481751}'s position matches with a bright and blue \gaia\ source. There is a hint of a very faint source to the south of this counterpart, but it should have minimal effect on the astrometric and photometric measurements. The counterpart is also a significant \galex\ UV source, which further favours the match. $\distance$ and $\dinv$ are consistent, giving a distance around $3.3\,\kpc$, and the \gaia\ photometry of J043509.3 is similar to that of J015648.6 and J025910.8, exhibiting a mild blue excess relative to the main sequence. Kinematically, J043509.3 is not particularly fast compared to other gold sources; indeed, the $\vpecminlo \approx 259\,\kms$ is only slightly above the selection limit. It has a non-zero $\aen$ of $0.16\,\mas$, but it is not significant at its distance.


{\bf 1eRASS J082815.9+383457} is the closest among the $\goldsamplesize$ gold sources, with a B21 inferred distance of only $\approx 600\,\pc$, consistent with its $\dinv$ ($\approx 1.7\,\mas$). The field around the X-ray position is also clear of other optical sources, making the \gaia\ counterpart a very confident match. Its $\bprp$ colour exhibits a clear blue excess relative to the main sequence, which is unlikely to be a chance coincidence. Our current sample of XRBs lacks such ``nearby” systems \citep[see e.g.,][]{Avakyan23, Neumann23}, and the discovery and confirmation of additional candidates would improve constraints on their local space density. 


{\bf 1eRASS J142449.5$-$153938}'s \gaia\ counterpart lies approximately $0.9\rerrx$ north-west of the X-ray position, and there is also a very faint source slightly to the east of the \gaia\ source, but it lies beyond $\rerrx$. The \gaia\ counterpart positionally coincides with a \galex\ UV source, so we consider it a genuine match to J142449.5. Both $\distance$ and $\dinv$ give an estimated distance around $1.2\,\kpc$, so its significant PM translates to a robust $\vpecminlo \approx 285\,\kms$. At this distance, the $\bprp$ colour is in line with the UV emission, showing an apparent blue excess relative to the bulk of the HVXS sample.

{\bf 1eRASS J152027.0$-$030609} is in a relatively uncrowded region, and its \gaia\ counterpart is the only source within its $\rerrx$. The counterpart is also matched to a \galex\ UV source, so we consider it a confident match to J152027.0. B21 estimates a $\distance\approx 3.8\,\kpc$, which is consistent with the $\dinv$ ($\approx 4.2\,\kpc$). J152027.0 is located on the red side of the upper main sequence, apparently separated from other gold sample sources.

\subsection{Uncertainties and limitations}
There are uncertainties that we need to be aware of when interpreting the HVXSs regarding their possible nature as COBs, and even their conservatively selected high space velocities. We address these issues in this section.

\subsubsection{Residual extragalactic contamination}
The first aspect to consider is residual extragalactic contaminants. AGNs and galaxies were primarily removed using the \gaia\ DCS classification (Sect \ref{sec:candidates_selection}), which achieves a relatively high completeness \citep[$\gtrsim 90\%$;][]{Delchambre23}, relying solely on \gaia's (optical) astrometry, photometry and (less commonly) spectroscopy. AGNs are X-ray emitters that meet the high-$\fxfg$ criterion, and the high-PM sources selected by the $\vpecminlo$ limit can also be mimicked by jet motion \citep[e.g.,][]{Khamitov23, Shen21}. To give a quantitative sense of the contamination level, we refer to \citet{Seppi22}, who estimated $\approx 64\%$ of the \erass\ sources are AGNs at its depth, while the DCS probability threshold (Sect \ref{sec:getting-a-kinematics-worthy-sample}) removed $\approx 33\%$ of the sources in the \erass\ catalogue. From another perspective, our crossmatch against SIMBAD suggests $\approx 37\%$ (out of $\numhvxsimbadmatches$ matches) of extragalactic contaminants, and this fraction translates to $\approx 878$ contaminants amongst the HVXS sample. Therefore, some AGNs likely slip through our selection. Future follow-up observations will refine the classifications.

\subsubsection{Uncertainty in source origin and kinematic interpretation}
\begin{figure}
    \centering
    \includegraphics[width=\columnwidth]{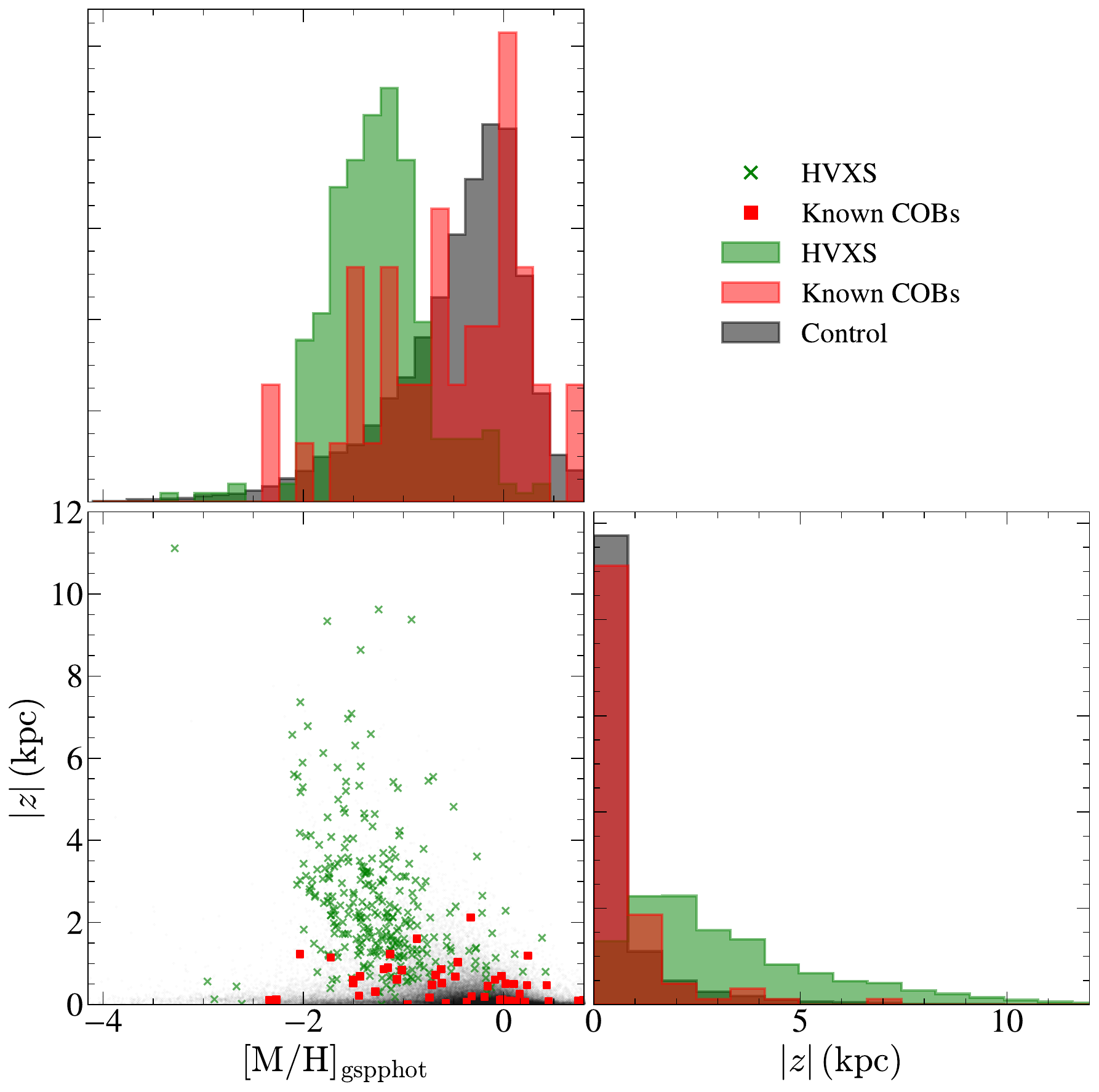}
    \caption{Vertical separation from the Galactic plane ($|z|$) plotted against the \gaia\ GSP-Phot metallicities ($\mhgspphot$) for the control (grey scale) and HVXS samples (green cross), as well as a subset of known COBs that have $\mhgspphot$ values (filled red squares). We use the median of $|z|$ and the nominal estimates of $\mhgspphot$ for the plots. HVXSs exhibits a broader distribution of $|z|$ and are generally more metal poor than the control sample.}
    \label{fig:mh_vs_z}
\end{figure}

Another key caveat is the formulation of $\vpecmin$. While subtracting rotation velocity effectively removes the contribution from bulk disc motion, it implicitly assumes a disc origin for all sources. Our cleaning procedure only excludes sources currently in clusters, while a more comprehensive kinematic determination of source origin would require constrained $\gamma$ measurements, which are beyond the scope of this work. There are indications, however, that some HVXSs may not originate from the disc. First, they appear kinematically heated, exhibiting a broader distribution in vertical distance from the Galactic plane ($|z|$) than the control sample, with only $\approx 6\%$ of HVXSs within $1\,\kpc$ of the Galactic plane. Second, they have generally lower metallicities than the control sample. We examine this by using the \gaia\ GSP-Phot metallicity estimates ($\mhgspphot$\footnote{All $\mhgspphot$ values have been calibrated using the {\tt gdr3apcal} package: \url{https://github.com/mpi-astronomy/gdr3apcal}.}). This is only available for a small ($10\%$) subset of HVXSs, but the distribution does peak around a sub-solar value of $-1.1$, with just $\approx 6\%$ above $-0.5$ --- the approximate median of $\mhgspphot$ for known COBs (Sect \ref{sec:where-to-draw-the-line}). We show the $\mhgspphot$ distributions of the HVXS, control, and known COB samples in Fig.~\ref{fig:mh_vs_z}. 

NKs are impulsive and energetic events, and when favourably oriented, they can propel disc-born COBs into the halo. This has been observed for some XRBs \citep[e.g.,][]{GonzalezHernandez08a, Leahy14} and BPSRs \citep[e.g.,][]{Shahbaz22}. For example, if a system orbiting at $238\,\kms$ in the solar neighbourhood (i.e., in the disc and $8.34\,\kpc$ away from the Galactic centre) received an instantaneous acceleration that increases its vertical velocity component to $200\,\kms$, it can reach $|z|_\mathrm{max}\approx 8\,\kpc$ in $\approx 80\,\mathrm{Myr}$ (as computed by {\sc galpy}). The object can reach an even higher $|z|$ at a greater distance, where it experienced less deceleration in the $z$ direction. However, NKs are not the only mechanism capable of producing such high velocities and large $|z|$. Some of these sources may instead be genuine halo objects that originated through other pathways. For example, halo stars and binaries can be formed in nearby satellite galaxies and then merged with the Milky Way, exhibiting distinct kinematic and chemical features. Alternatively, high --- and even runaway --- velocities can be produced by dynamical processes in dense environments, such as globular clusters \citep[e.g.,][]{vanParadijs95, Irwin05, Cabrera23} and the Galactic bulge, or close encounter with single \citep{Hills88} or binary \citep{Yu03} massive black holes. Our search is intended to be inclusive and comprehensive, but it would greatly benefit from large-scale metallicity surveys, particularly for faint sources that are understudied in this regard.

\subsection{Selection completeness and trade-offs}
We began with a sizeable sample of X-ray sources, which was reduced dramatically through successive selection steps --- from over 2 million sources to $\numhvxaftercleaning$ HVXSs. The most stringent quality cut retained only $\goldsamplesize$ sources in our `gold' sample. Within the scope of existing source catalogues, two main steps contributed most to this reduction: (1) the application of a $\vpecminlo$ threshold combined with the $\fxfg$ cut (Fig. \ref{fig:flow-chart-hvxs}), and (2) the deduplication of matches and exclusion of matches beyond $2\,\rerrx$ (Sect \ref{sec:candidates_selection}).Due to the large uncertainties in distance estimates, many sources have correspondingly large errors on $\vpecmin$. As a result, applying a $1\,\sigma$ lower limit ($\vpecminlo$) excludes marginally fast sources with $\vpecmin \approx \vpecminlim\,\kms$. In addition, criterion (2) removes sources located in crowded fields (e.g., the Galactic bulge), where multiple optical matches are common. 

On the other hand, disc-born high-velocity systems may be inherently rare --- not only because a considerable fraction are ejected, but also because those that remain bound can migrate to large distances from the disc and spend longer times there, making them harder to detect in flux-limited surveys. Beyond the current catalogues, many sky regions are yet to be covered beyond shallow depths. Future releases of eRASS with full-sky coverage, together with deeper astrometric surveys (upcoming \gaia\ DR4 and future surveys with the Roman telescope and others), will enable us to push deeper and circumvent our selection biases. Ultimately, these should help to better understand the selection function and extract underlying population densities of the fast-moving X-ray source population.

\section{Conclusions}
\label{sec:conclusions}
We have compared the distributions of $\vpec$ of known compact object binaries (COBs) with that of major contaminants and found that $\vpecminlim\,\kms$ can exclude over $97\%$ of contaminants. Adopting this limit on the $1\,\sigma$ lower limit of $\vpecmin$, we performed a comprehensive search for HVXSs that exhibit $\fxfg$ ratios clearly over the empirical limit found by R24. We found a total of $\numhvxaftercleaning$ sources with confident counterparts, and we curated a gold sample of $\goldsamplesize$ sources. These X-ray sources may originate from COBs and can be followed-up for features of interaction (e.g., accretion). Metallicity constraints from future large-scale spectroscopic surveys will further distinguish their origins.
\section*{Acknowledgements}
We thank the anonymous reviewer for their helpful comments. PG is a Royal Society Senior Leverhulme Trust fellow, and also thanks STFC for support. PC acknowledges the Leverhulme Trust for an Emeritus Fellowship. The work of DS was carried out at the Jet Propulsion Laboratory, California Institute of Technology, under a contract with the National Aeronautics and Space Administration (80NM0018D0004).

This work has made use of data and resources from the following: 
The European Space Agency (ESA) mission {\it Gaia} (\url{https://www.cosmos.esa.int/gaia}), processed by the {\it Gaia} Data Processing and Analysis Consortium (DPAC, \url{https://www.cosmos.esa.int/web/gaia/dpac/consortium}). Funding for the DPAC
has been provided by national institutions, in particular the institutions
participating in the {\it Gaia} Multilateral Agreement;
the cross-match service provided by CDS, Strasbourg;
the SIMBAD database, operated at CDS, Strasbourg, France;
the Aladin Sky Atlas, CDS, Strasbourg Astronomical Observatory, France.
X-ray source catalogues are from multiple resources, including data obtained from the Chandra Source Catalog, provided by the Chandra X-ray Center (CXC), the 4XMM XMM-Newton serendipitous source catalogue compiled by the XMM-Newton Survey Science Centre, the Swift XRT Point Source catalogue obtained from the UK Swift Science Data Centre at the University of Leicester, and the eROSITA all-sky survey catalogue. eROSITA is the soft X-ray instrument aboard SRG, a joint Russian-German science mission supported by the Russian Space Agency (Roskosmos), in the interests of the Russian Academy of Sciences represented by its Space Research Institute (IKI), and the Deutsches Zentrum für Luft- und Raumfahrt (DLR). The SRG spacecraft was built by Lavochkin Association (NPOL) and its subcontractors, and is operated by NPOL with support from the Max Planck Institute for Extraterrestrial Physics (MPE). The development and construction of the eROSITA X-ray instrument was led by MPE, with contributions from the Dr. Karl Remeis Observatory Bamberg \& ECAP (FAU Erlangen-Nuernberg), the University of Hamburg Observatory, the Leibniz Institute for Astrophysics Potsdam (AIP), and the Institute for Astronomy and Astrophysics of the University of Tübingen, with the support of DLR and the Max Planck Society. The Argelander Institute for Astronomy of the University of Bonn and the Ludwig Maximilians Universität Munich also participated in the science preparation for eROSITA. 

This work also made use of images from the SkyMapper Southern Sky Survey. The national facility capability for SkyMapper has been funded through ARC LIEF grant LE130100104 from the Australian Research Council, awarded to the University of Sydney, the Australian National University, Swinburne University of Technology, the University of Queensland, the University of Western Australia, the University of Melbourne, Curtin University of Technology, Monash University and the Australian Astronomical Observatory. SkyMapper is owned and operated by The Australian National University's Research School of Astronomy and Astrophysics. The survey data were processed and provided by the SkyMapper Team at ANU. The SkyMapper node of the All-Sky Virtual Observatory (ASVO) is hosted at the National Computational Infrastructure (NCI). Development and support of the SkyMapper node of the ASVO has been funded in part by Astronomy Australia Limited (AAL) and the Australian Government through the Commonwealth's Education Investment Fund (EIF) and National Collaborative Research Infrastructure Strategy (NCRIS), particularly the National eResearch Collaboration Tools and Resources (NeCTAR) and the Australian National Data Service Projects (ANDS)

In addition to the software mentioned in the texts, analysis and visualisation in this work have also made use of the following packages (in alphabetic order): {\sc astropy}:\footnote{http://www.astropy.org} a community-developed core Python package and an ecosystem of tools and resources for astronomy \citep{Astropy22}, {\sc matplotlib} \citep{Matplotlib07}, {\sc NetworkX} \citep{networkx08}, {\sc numpy} \citep{Numpy20}, {\sc pandas} \citep{Pandas23}, {\sc scipy} \citep{Scipy20}, and {\sc topcat} \citep{Taylor05}.
\section*{Data Availability}
The X-ray source catalogues used in this work are all publicly available and has been updated regularly. The \chandra\ Source Catalogue 2.1 can be accessed and queried using the CSCview application\footnote{\url{https://cda.cfa.harvard.edu/cscview/}}, the {\it XMM-Newton} Serendipitous Source Catalogue can be downloaded directly from the {\it XMM-Newton} Survey Science Center\footnote{\url{https://xmmssc.aip.de/cms/catalogues/4xmm-dr14s/}}, the \swift\ Point Source Catalogue can be downloaded from the official website\footnote{\url{https://www.swift.ac.uk/2SXPS/}}, and finally, the \erass\ DE DR1 source catalogue can be accessed and downloaded from the DR1 website\footnote{\url{https://erosita.mpe.mpg.de/dr1/AllSkySurveyData_dr1/Catalogues_dr1/}}. All \gaia-related catalogues can be accessed through the \gaia\ archive\footnote{https://gea.esac.esa.int/archive/}.




\bibliographystyle{mnras}
\bibliography{refs} 



\appendix

\section{Additional tables}

\subsection{Kept SIMBAD types}
\label{sec:removed-and-kept-simbad-types}

Table \ref{tab:kept-simbad-types} lists SIMBAD types kept in our selection (Sect \ref{sec:candidates_selection}).

\begin{table*}
    \centering
    \caption{Kept SIMBAD types}
    \begin{tabular}{ll|ll|ll}
    \hline
    Type & Counts & Type & Counts & Type & Counts \\
    \hline
    Radio        & 41  & cmRad      & 2 & Transient & 1 \\
    Star         & 37  & RotV*      & 1 & WhiteDwarf\_Candidate & 1  \\
    X            & 19  & Variable*  & 1 &  & \\
    EclBin       & 9   & Eruptive*  & 1 &  & \\
    HighPM*      & 6   & ChemPec*   & 1 &  & \\
    LongPeriodV* & 2   & Low-Mass*  & 1 &  & \\
    gammaBurst   & 2   & blue       & 1 &  & \\
    \hline
    \end{tabular}
    \label{tab:kept-simbad-types}
\end{table*}


\subsection{Acronyms and symbols}
This section lists acronyms (Table \ref{tab:acronyms}) and symbols (Table \ref{tab:symbols}) defined throughout the texts.

\begin{table}
\centering
    \caption{Acronyms (in alphabetic order) used in this work}
    \begin{tabularx}{\columnwidth}{lX}
        \hline
        \textbf{Acronym} & \textbf{Definition} \\
        \hline
        AGN & Active galactic nucleus \\
        BH & Black hole \\
        BPSR & Binary pulsar \\
        COB & Compact object binary \\
        CSC & \chandra\ Source Catalogue \\
        CV & Cataclysmic variable \\
        eRASS & eROSITA All-Sky Survey \\
        GSP-Phot & General Stellar Parametrizer from Photometry module \\
        HMXB & High-mass X-ray binary \\
        HVXS & High-velocity X-ray source\\
        LMC & Large Magellanic Cloud \\
        LMXB & Low-mass X-ray binary \\
        LSR & Local standard of rest \\
        NICOB & Non-interacting compact object binary \\
        NK & Natal kick \\
        NS & Neutron star \\
        PM & Proper motion \\
        RV & Radial velocity \\
        SMC & Small Magellanic Cloud \\
        XRB & X-ray binary \\
        YSO & Young stellar object \\
        \hline
    \end{tabularx}
\label{tab:acronyms}
\end{table}

\begin{table}
\centering
    \caption{Symbols defined in this work}
    \begin{tabularx}{\columnwidth}{lX}
        \hline
        \textbf{Symbol} & \textbf{Definition} \\
        \hline
        $\ra$ & Right ascension \\
        $\dec$ & Declination \\
        $\pmracosdec$ & PM component in the direction of $\ra$ \\
        $\pmdec{}$ & PM component in the direction of $\dec$. \\
        $\parallax$ & Parallax \\
        $\parallaxerror$ & Uncertainty in $\parallax$ \\
        $\distance$ & Distance from \citet{Bailer-Jones21} \\
        $\dinv$ & Distance from inverting $\parallax$ \\
        $\aen$ & Astrometric excess noise \\
        $\aensig$ & Significance of $\aen$ \\
        $\vpec$ & Peculiar velocity relative to Galactic rotation \\
        $\vpecmin$ & Minimum peculiar velocity \\
        $\vpecminlo$ & $1\,\sigma$ lower limit of $\vpecmin$ \\
        $\gamma$ & Systemic radial velocity \\
        $\vpecgammamin$ & $\gamma$ value that minimises $\vpec$ \\
        $\gaiaxsep$ & Angular separation of \gaia\ counterpart to X-ray position \\
        $\rerrx$ & Uncertainty in X-ray position \\
        $\gabs$ & Absolute \gaia\ G-band magnitude \\
        $\gmag$ & Apparent \gaia\ G-band magnitude \\
        $\bprp$ & \gaia\ Bp$-$Rp colour \\
        $\fxfg$ & X-ray/\gaia\ G flux ratio \\
        $\mhgspphot$ & \gaia\ GSP-Phot metallicity \\
        $|z|$ & Vertical separation from the Galactic plane \\
        \hline
    \end{tabularx}
\label{tab:symbols}
\end{table}

\bsp	
\label{lastpage}
\end{document}